%% file: recovery_time_paper.tex
\documentclass[journal,twoside,onecolumn,letterpaper]{IEEEtran}
\usepackage{graphicx}
\usepackage{dblfloatfix}    % To enable figures at the bottom of page
\usepackage[monochrome]{xcolor}
\usepackage{psfrag}
\usepackage[caption=false,font=footnotesize]{subfig}

\usepackage{multirow}
\usepackage{fixltx2e}
\usepackage[cmex10]{amsmath}
\usepackage{amsfonts}
\usepackage{amssymb}
\usepackage{amsmath}
\usepackage{verbatim}
\usepackage{arydshln}

\usepackage{tikz}
\usepackage[numbers, sort]{natbib}
\usetikzlibrary{arrows,automata}
\tikzset{every state/.style={minimum size=0pt}}
\usepackage[acronym,nonumberlist,sort=use]{glossaries}
\input{myglossary}
\makeglossaries

\newcommand{\cW}{\mathcal{W}} %%% window of suscpetibility
\newcommand{\cA}{\mathcal{A}} %%% window of suscpetibility
\newcommand{\cB}{\mathcal{B}} %%% window of suscpetibility
\newcommand{\cC}{\mathcal{C}} %%% window of suscpetibility
\newcommand{\cD}{\mathcal{D}} %%% window of suscpetibility
\newcommand{\cX}{\mathcal{X}} %%% window of suscpetibility

\newcommand{\prob}{\text{Pr}}

\makeatletter
\renewcommand*\env@matrix[1][*\c@MaxMatrixCols c]{%
  \hskip -\arraycolsep
  \let\@ifnextchar\new@ifnextchar
  \array{#1}}
\makeatother
\renewcommand*{\arraystretch}{1.3}

\begin{document}
\title{
Effect of Jitter on the Settling Time of Mesochronous Clock Retiming Circuits
%Settling Time of Mesochronous Clock Retiming Circuits in the Presence of 
%Timing Jitter
}

\author{Naveen Kadayinti$^\text{1}$, 
Amitalok J. Budkuley$^\text{2}$, 
Maryam S. Baghini$^\text{3}$, 
and Dinesh K. Sharma$^\text{4}$
\thanks{This paper was presented in part at ISCAS 2017, Baltimore, USA, 
May 28-31, 2017.

}

\thanks{
$^\text{1}$ The Department of Electrical Engineering, Indian 
Institute of Technology Dharwad, Dharwad, Karnataka, India 

$^\text{2}$ Department of Information Engineering, The Chinese
University of Hong Kong, Sha Tin, Hong Kong 

$^\text{3,4}$ The Department of Electrical Engineering, Indian Institute of
Technology Bombay, Mumbai, India

}

\thanks{
This work was supported by the Tata Consultancy Services (TCS)
and the SMDP programme of the Government of India, in the form of
student scholarships and CAD tool licenses respectively.}
}

\maketitle
\input{abstract}

\begin{IEEEkeywords}
Settling time, clock recovery, metastability, inter-symbol interference, 
low swing interconnect, absorbing Markov chains.
%training sequence design.
\end{IEEEkeywords}

\input{./intro.tex}

\input{./settling.tex}
\input{./sync.tex}
\input{./markov.tex}

\input{./quantification.tex}

\input{./reduction.tex}
\input{./conclusions.tex}

\input{./appendix.tex}

\bibliographystyle{IEEE}
\bibliography{IEEEabrv,refs}
\input{bio}
\end{document}

%% file: myglossary.tex
\newacronym{cdf}{CDF}{cumulative distribution function}
\newacronym{dll}{DLL}{delay locked loop}
\newacronym{pll}{PLL}{Phase Locked Loop}
\newacronym{bist}{BIST}{Built In Self Test}
\newacronym{lpf}{LPF}{Low Pass Filter}
\newacronym{cdr}{CDR}{clock and data recovery}
\newacronym{ccffe}{CCFFE}{Capacitively Coupled Feed Forward Equalizer}
\newacronym{ffe}{FFE}{Feed Forward Equalizer}
\newacronym{ber}{BER}{Bit Error Rate}
\newacronym{wcc}{WCC}{Worst Case Condition}
\newacronym{ui}{UI}{Unit Interval}
\newacronym{isi}{ISI}{inter-symbol-interference}
\newacronym{mtbf}{MTBF}{Mean Time Between Failures}
\newacronym{dfe}{DFE}{Decision Feedback Equalizer}
\newacronym{CMRR}{CMRR}{Common Mode Rejection Ratio}
\newacronym{vcdl}{VCDL}{Voltage Controlled Delay Line}
\newacronym{prbs}{PRBS}{Pseudo Random Binary Sequence Generator}
\newacronym{fsm}{FSM}{finite state machine}
\newacronym{cad}{CAD}{Computer Aided Design}
\newacronym{dvfs}{DVFS}{Dynamic Voltage and Frequency Scaling}
\newacronym{adc}{ADC}{Analog to Digital Converter}
\newacronym{dac}{DAC}{Digital to Analog Converter}
\newacronym{fir}{FIR}{Finite Impulse Response}
\newacronym{tdc}{TDC}{Time to Digital Converter}
\newacronym{dtc}{DTC}{Digital to Time Converter}
\newacronym{enob}{ENOB}{Effective Number Of Bits}
\newacronym{osr}{OSR}{Over Sampling Ratio}
\newacronym{fpga}{FPGA}{Field Programmable Gate Array}

\newglossaryentry{ff}{
name=FF,
description = Fast N and Fast P corner}

\newglossaryentry{ss}{
name=SS,
description = Slow N and Slow P corner}

\newglossaryentry{fs}{
name=FNSP,
description = Fast N and Slow P corner}

\newglossaryentry{sf}{
name=SNFP,
description=Slow N and Fast P corner}

\newglossaryentry{tt}{
name=TT,
description = Typical process parameters}

\newglossaryentry{vth}{
name=$V_{th}$,
description = Threshold voltage of the transistor
}

\newglossaryentry{pi}{
name=$\pi$,
description = Ratio of circumference to diameter of a circle(3.142)
}
\newglossaryentry{epsilon}{
name=$\epsilon _0$,
description = Permittivity of free space ($8.85\times 10^{-12}F/m$)
}

\newglossaryentry{cktx}{
name=$\phi_{Tx}$,
description = Phase of the transmitter clock}

\newglossaryentry{ckrx}{
name=$\phi_{Rx}$,
description = Phase of the receiver clock}

\newglossaryentry{ckd}{
name=$\phi_{d}$,
description = Phase of the sampling clock}

\newglossaryentry{cW}{
name=$\mathcal{W}$,
description = Window of susceptibility to extended settling time}

\newglossaryentry{TcW}{
name=$T_\mathcal{W}$,
description = Width of the window of susceptibility $\mathcal{W}$}

\newglossaryentry{cke}{
name=$\Delta \phi$,%e^{0}$,
description = Initial phase error between the data and the sampling clock}

\newglossaryentry{ckd0}{
name=$\phi_d^0$,
description = Initial value of the sampling clock phase}

\newglossaryentry{dact}{
name=$\alpha$,
description = Data activity factor}

%% file: abstract.tex
\begin{abstract}
It is well known that timing jitter can degrade the bit error rate 
(BER) of receivers that recover the clock from input data. However, 
timing jitter can also result in an indefinite increase in the 
settling time of clock recovery circuits, particularly in low swing 
mesochronous systems. Mesochronous clock retiming circuits are 
required in repeaterless low swing on-chip interconnects. We first 
discuss how timing jitter can result in a large increase in the 
settling time of the clock recovery circuit. Next, the circuit is 
modelled as a Markov chain with absorbing states. The mean time to 
absorption of the Markov chain, which represents the mean settling 
time of the circuit, is determined. The model is validated through 
behavioural simulations of the circuit, the results of which match 
well with the model predictions. We consider circuits with 
(i)~data dependent jitter, (ii)~random jitter, and (iii) combination 
of both of them. We show that a mismatch between the strengths 
of up and down corrections of the retiming can reduce the settling 
time. In particular, a 10\% mismatch can reduce the mean settling 
time by up to 40\%. 
We leverage this fact toward  improving the settling time 
performance, and propose useful techniques based on biased training 
sequences and  mismatched charge pumps. 
%This mismatch can be implemented through a 
%training sequence or by incorporating this mismatch in the charge 
%pump of the retiming circuit and both these techniques are discussed 
%in this paper. 
We also present a coarse+fine clock retiming circuit, 
which can operate in coarse first mode, to reduce the settling time 
substantially. These fast settling retiming circuits are verified 
with circuit simulations.

\end{abstract}

%% file: intro.tex
\section{Introduction}
\label{sec:intro}
Timing jitter in the incoming data degrades the bit error rate (BER)
of receivers that recover clock from the data itself, and this effect 
has been extensively studied (c.f.,~\cite{CDR_razavi,jitter-ber}). 
However, it has been recently observed in~\cite{naveen_iscas17} that, 
under certain conditions, timing jitter can also increase the settling 
time of clock retiming circuits. In this paper, we present a detailed 
analysis of the settling time of mesochronous clock retiming circuits 
in the presence of jitter. 

%Timing jitter can 
%also increase the settling time of clock retiming circuits in 
%certain conditions~\cite{naveen_iscas17}. This dependence of the 
%settling time on timing jitter has not been analyzed well till now. 
%In this paper, we analyze the settling time of mesochronous clock 
%retiming circuits in the presence of different types of jitter.
%However, the dependence 
%of settling time of clock retiming circuits on the jitter in the 
%incoming data has not been investigated before. This paper discusses 
%the effect of timing jitter on the settling time of mesochronous 
%clock retiming circuits.

Mesochronous clock retiming circuits are required in repeaterless low 
swing on-chip interconnects to sample the data correctly at the 
receiver~\cite{Lee-jssc14,naveen_vlsi17}, and also in off-chip links 
that use a forwarded clock \cite{fwd_clock,fwd_clock2}. In mesochronous 
receivers, a clock running at the correct frequency is available at 
the receiver and only the correct phase needs to be recovered. Hence,
such clock retiming circuits use delay 
line~\cite{naveen_vlsi17,Lee-jssc14} or phase 
interpolators~\cite{phase_interpolator} to generate the required clock 
phase. Use of delay based retiming circuits is preferred over systems 
that use phase locked loops with a voltage controlled oscillator (VCO) 
per channel due to their better performance and lower 
complexity~\cite{dig_pi_jssc03}. Delay based retiming circuits are 
preferred even in systems where the clock frequency is known only 
nominally and not exactly~\cite{phase_interpolator}.
Fast locking clock data recovery (CDR) circuits (also known as 
burst mode CDRs~\cite{oversampling-cdr}) offer very short settling 
times. 
% they suffer from certain limitations. 
These CDR circuits use gated VCOs~\cite{gvco-cdr}, 
oversampling~\cite{oversampling-cdr} or 
injection locking~\cite{burst-cdr-inj}. However, these CDR circuits 
have certain limitations. Gated VCOs have poor phase aligning and 
no input jitter rejection. Oversampling CDRs come with high complexity 
and power consumption. Injection locked CDRs have quantization phase 
error and high cycle to cycle jitter. Hence, for source synchronous 
on-chip and chip to chip links, delay based retiming circuits are 
preferred. Hsieh et al., in~\cite{cdr-review}, provide a detailed 
discussion and comparison of different CDR circuit architectures.

Delay based clock retiming circuits are known to offer fast settling times 
as compared to 
PLL based synchronizers as they do not need a clock synthesis 
step~\cite{cdr-review}. 
Timing jitter can increase the settling time of such delay based clock
retiming circuits indefinitely\cite{naveen_iscas17}.
This occurs when the circuit wakes up with its clock in the horizontally closed 
region of the data eye. Typical systems implemented on-chip can have hundreds 
of long interconnects~\cite{global_wires,wire-length}, and a horizontal eye 
opening of about 85\% is not pessimistic~\cite{Mensink-jssc-2010,kim-jssc2010}. 
If the initial clock position is assumed to be uniformly 
distributed, there is a 15\% chance that the circuit wakes up 
with its initial clock phase in the closed region of the eye, making it 
important to study and understand this problem. 
To analyze the settling time, the clock retiming circuit can be modelled as 
a Markov chain with absorbing states~\cite{naveen_iscas17}. Here 
the states correspond to the 
clock positions. The state transitions correspond to the 
phase corrections done by the clock recovery circuit. 
This model provides useful insights into the dynamics of the system. 
In particular, as shown in this work, 
the mean settling time of the circuit is predicted by the model.

In this paper, we investigate the settling time of mesochronous clock 
retiming circuits in detail. Section~\ref{sec:settling} presents a 
detailed discussion on the settling time of mesochronous clock retiming 
circuits, in particular, its dependence on timing jitter, which is the 
main focus of this work. We demonstrate the increase in settling time, 
caused due to jitter, using experiments performed on a chip and these 
experiments are described 
%An experimental study of this degradation of 
%the settling time is given 
in Section~\ref{sec:expt}. Next, toward 
analyzing different types of jitter, a Markov chain model of the 
mesochronous synchronizers is presented in Section~\ref{sec:markov}. 
As the settling time is random, it cannot be determined exactly. 
However, using the Markov chain model we can determine bounds on the 
settling time. These are given in Section~\ref{sec:quant}. Techniques 
for reducing the settling time, along with supporting circuit 
simulations, are given in Section~\ref{sec:reduce}. Finally, we make 
overall concluding remarks in Section~\ref{sec:conclude}.

%% file: settling.tex
\section{Settling time of the clock retiming circuit}
\label{sec:settling}
\begin{figure}[!h]
\psfrag{CkTx}{\small{\gls{cktx}}}
\psfrag{CkRx}{\small{\gls{ckrx}}}
\psfrag{n+alpha}{\small{$(n+\lambda)\times T$}}
\psfrag{Ckd}{\small{\gls{ckd}}}
\psfrag{Source}{\small{Source}}
\psfrag{Sense}{\small{Sense}}
\psfrag{Amplifier}{\small{Amplifier}}
\psfrag{Destination}{\small{Destination}}
\psfrag{Retiming}{\small{Clock Retimer}}
\psfrag{Line}{\small{Interconnect}}
\centering
\includegraphics[width=0.47\columnwidth]{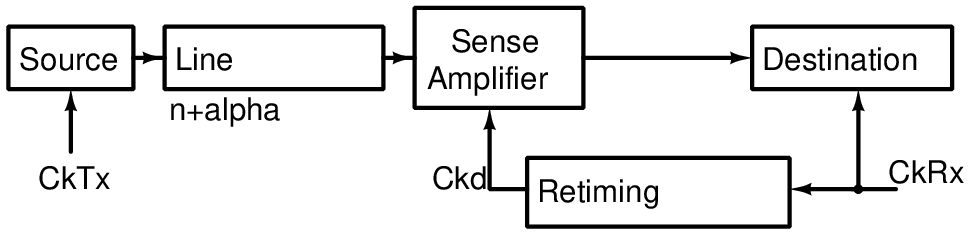}
\caption[Block diagram of a repeaterless low
swing interconnect system.]{Block diagram of a repeaterless low
swing interconnect system. \mbox{\gls{cktx}: transmitter clock}, 
\mbox{\gls{ckrx}}: 
receiver clock, \mbox{\gls{ckd}}: retimed sampling clock, 
($(n+\lambda)\times T$): repeaterless interconnect delay, where
\mbox{$n \in \mathbb{Z}^{+}$},
\mbox{$\lambda \in [0,1)$,
%\mathbb{R}$} \& \mbox{$\lambda < 1$}, 
\mbox{$T$: system clock period.}}}
\label{fig:ch6:interconnect_system}
\end{figure}

\begin{figure}[h]
\centering
\psfrag{w}{\scriptsize{\hspace{0.5ex}$\cW$}}
\psfrag{0}{\small{$0$}}
\psfrag{2pi}{\small{$2\pi$}}
\psfrag{ck}{\small{$\phi_d^0$}}
\includegraphics[width=4cm]{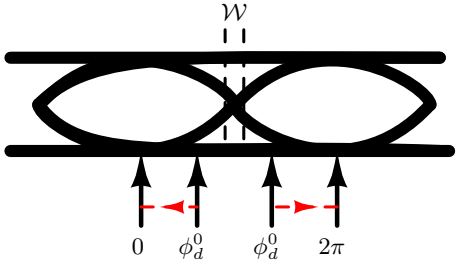}
\caption{Illustration of a data eye diagram indicating the sampling 
clock's initial position along with direction of phase corrections for 
two cases, i.e., $\phi_d^0 > \pi$ and $\phi_d^0 < \pi$.}
\label{fig:ch6:eye-cartoon2}
\end{figure}

In order to analyze the settling time of a mesochronous system, we 
consider a repeaterless interconnect system as shown in 
Fig.~\ref{fig:ch6:interconnect_system}. Here the delay of the 
interconnect is expressed as $(n+\lambda)T$, where 
$n \in \mathbb{Z}^{+}$, $\lambda \in [0,1)$ and $T$ is the system clock 
period.
\gls{cktx} and \gls{ckrx} are the phases of the transmitter and 
receiver clocks respectively. The clock retiming circuit derives the 
sampling clock \gls{ckd}, which  is positioned at the center of the 
input data eye, from the receiver clock phase. The settling time of 
the circuit is defined as the time it takes the circuit to derive this 
sampling clock phase. This time depends on the initial phase error 
between the clock and the data. This initial phase error \gls{cke} is 
a continuous variable taking values in $[0, 2\pi)$. When the initial 
phase error is less than $\pi$, the circuit achieves lock by decreasing 
the phase to 0. On the other hand, when \gls{cke} is greater than 
$\pi$, the circuit achieves lock by increasing the phase difference to 
$2\pi$, which is the same as 0 by phase wrapping (refer 
Fig.~\ref{fig:ch6:eye-cartoon2}).
The settling time of the clock recovery circuit depends on the gain of 
the system and the initial phase error. For mesochronous systems, this 
settling time can be expressed as
\begin{align}
\label{eqn:ch6:ts_deterministic}
t_s=
\begin{cases}
T\frac{C}{\alpha K_{VC}K_{CP}} \Delta \phi &\mbox{ when } \Delta  \phi < \pi ,\\ 
T\frac{C}{\alpha K_{VC}K_{CP}}(2\pi - \Delta \phi) &\mbox{ when } \Delta  \phi > \pi .
\end{cases}
\end{align}
Here $C$ is the loop filter capacitor, $K_{VC}$ is the gain of the phase 
modulator, $K_{CP}$ is the gain of the charge pump and $\alpha$ is the 
data activity factor. The expression in \eqref{eqn:ch6:ts_deterministic} 
is derived in Appendix~\ref{apdx:ts}.

\subsection{Lengthening of the settling time}
When the initial phase difference is equal to $\pi$, the circuit can 
settle at two discrete values of the phase difference: $0$ or $2\pi$ 
(which is same as $0$ by phase wrapping), and both the solutions are 
acceptable. Since the initial phase difference is a continuous variable, the above scenario is one of taking a discrete decision on a continuous 
input. Theoretically, such decisions can take an infinite amount of 
time for certain initial conditions~\cite{lamport}. This happens when 
\gls{cke} is exactly $\pi$ radians and the system has no reason to 
choose one solution over another. This is akin to metastability in 
flip-flops~\cite{metastability-chaney}. When \gls{cke} is exactly 
$\pi$ radians, the flip flops in the phase detector become metastable. 
In these conditions, the phase detector loop enters a state of 
indecision. However, for sustained indecision of the phase detector 
loop, the flip-flops in the phase detector must become metastable in 
every clock cycle. Practically, however, given the narrow widths of the 
metastability windows, the fast recovery times and the inherent jitter 
present in the clocks, sustained loop indecision due to flip-flop 
metastability is highly unlikely. Hence, metastability of the 
flip-flops in the phase detector is \emph{not} discussed in this work.

\subsection{Effect of timing jitter: The \emph{window of susceptibility}}
\begin{figure}[h!]
\centering
\psfrag{w}{\scriptsize{\hspace{0.5ex}$\cW$}}
\psfrag{0}{\small{$0$}}
\psfrag{2pi}{\small{$2\pi$}}
\psfrag{ck}{\small{$\phi_d^0$}}
\includegraphics[width=4cm]{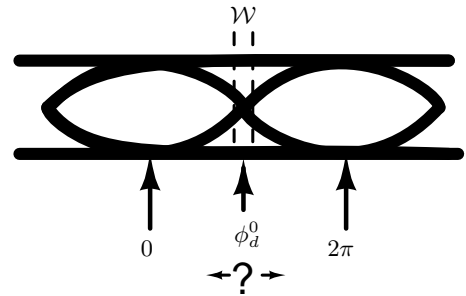}
\caption{Illustration of an eye diagram with timing jitter and 
with initial clock position $\phi_d^0$ in the horizontally closed region $\cW$.}
\label{fig:ch6:eye-cartoon}
\end{figure}

When the input data has timing jitter (due to \gls{isi} in the data 
and/or random jitter in the clock) and the initial clock phase 
(\gls{ckd0}) is in the horizontally closed region of the data eye, the 
expression of settling time in~\eqref{eqn:ch6:ts_deterministic} is not 
valid. Fig.~\ref{fig:ch6:eye-cartoon} illustrates an eye diagram with 
timing jitter and when \gls{ckd0} is in the horizontally closed region 
of the data eye. 
This region is called the \emph{window of susceptibility} (\gls{cW}) 
henceforth. When the initial clock is in this window $\mathcal{W,}$
the output of the phase detector is randomized by the jitter in the 
data, and hence, the phase error information is lost. This increases 
the settling time $t_s$ indefinitely, till the system escapes this 
window $\cW$. 

%% file: sync.tex
\subsection{Working of phase detectors in the window of
susceptibility}
\label{sec:manifestation}
\begin{figure}[h!]
\centering
\psfrag{Data}{\small{Data}}
\psfrag{Clock}{\small{Clock}}
\psfrag{Clock Early}{\small{Clock Early}}
\psfrag{Clock Late}{\small{Clock Late}}
\psfrag{A}{\small{A}}
\psfrag{B}{\small{B}}
\psfrag{C}{\small{C}}
\psfrag{(a)}{\small{(a)}}
\psfrag{(b)}{\small{(b)}}
\psfrag{(c)}{\small{(c)}}
\psfrag{Data}{\small{Data}}
\psfrag{Clock}{\small{Clock}}
\psfrag{A}{\small{A}}
\psfrag{B}{\small{B}}
\psfrag{C}{\small{C}}
\psfrag{UP}{\small{UP}}
\psfrag{DN}{\small{DN}}
\subfloat[Sampling instants of the phase detector]{
%\begin{subfigure}[b]{0.9\columnwidth}
%\centering
\includegraphics[height=3cm]{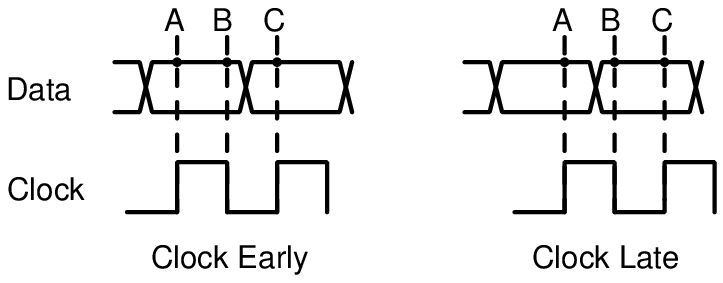}
\label{fig:ch6:alex_sampling_ideal}} \\
\subfloat[Sampling in the presence of jitter]{
\includegraphics[height=3.2cm]{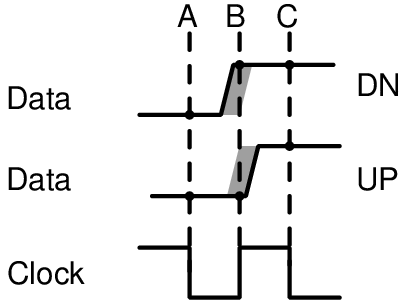}
\label{fig:ch6:alex_sampling_jitter}}
\caption{The sampling instants of the Alexander phase detector,
(a) under normal conditions, and (b) 
when $\Delta \phi \sim \pi$ and the data is corrupt with 
jitter~\cite{naveen_iscas17}.}
\label{fig:ch6:alex_timing_isi}
\end{figure}

The timing diagram of the Alexander phase detector~\cite{alexander_pd} 
is shown in 
Fig.~\ref{fig:ch6:alex_timing_isi}\subref{fig:ch6:alex_sampling_ideal}. 
The phase detector takes 2 samples per bit period and takes a binary 
decision of shifting the clock to the right or to the left based on 
the last three samples (labeled as A, B and C in 
Fig.~\ref{fig:ch6:alex_timing_isi}). When there is no jitter in the 
data, the phase detector consistently produces either up (UP) or 
down (DN) pulses. 
Thus, for a given data activity factor \gls{dact}, the settling time 
can be calculated using~\eqref{eqn:ch6:ts_deterministic}.

The phase detector's decision to assert one of UP or DN signals 
depends on the value of the sample taken at time instant B.
Fig.~\ref{fig:ch6:alex_timing_isi}\subref{fig:ch6:alex_sampling_jitter} 
shows the timing diagram of the Alexander phase detector when the data 
is corrupt with timing jitter and the initial phase error is close to 
$\pi$ radians, i.e., the initial clock position is in the window 
\gls{cW}~\cite{naveen_iscas17}. Due to jitter the value of the sample at time instant B depends on the 
current data transition time and not on the average data arrival time.
Hence, the phase detector produces UP and DN pulses randomly and the 
clock recovery circuit can remain stuck, with its clock in this region, 
indefinitely.

\begin{figure}[h!]
\centering
\psfrag{alex}{\footnotesize{Alexander PD}}
\psfrag{Hogge}{\footnotesize{Hogge PD}}
\psfrag{Tw}{\footnotesize{$T_{\cW}$}}
\psfrag{pi}{\footnotesize{$\pi$}}
\psfrag{2pi}{\footnotesize{$2\pi$}}
\psfrag{0}{\footnotesize{0}}
\psfrag{+1}{\footnotesize{$+1$}}
\psfrag{-1}{\footnotesize{$-1$}}
\includegraphics[width=0.385\columnwidth]{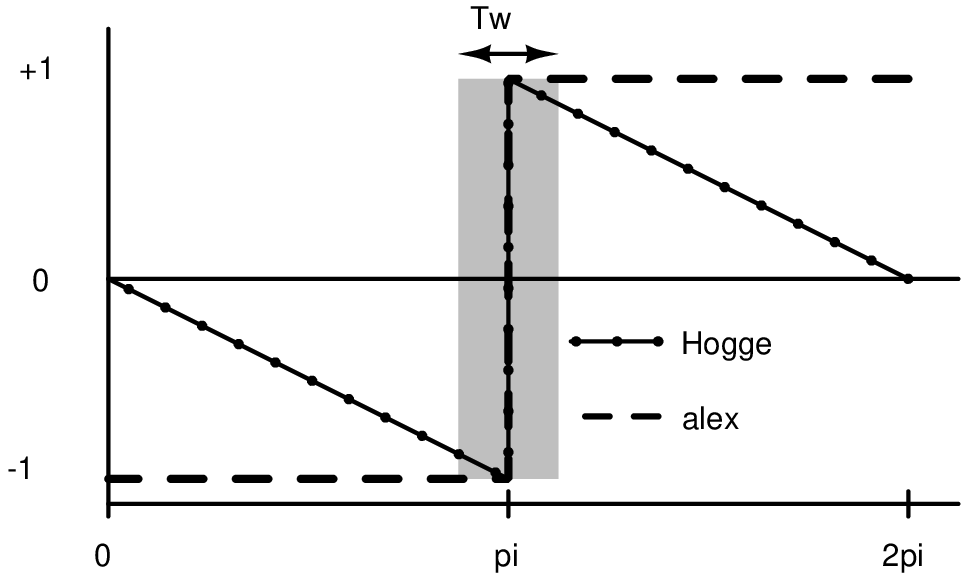}
\caption{The characteristics of linear and binary phase detectors (PD).
Highlighted region indicates the window $\cW.$ 
The phase detector gains are normalized.}
\label{fig:ch6:pd-chars}
\end{figure}

\begin{figure}[h!]
\centering
\psfrag{Tw1}{\small{$T_{\cW}^1$}}
\psfrag{Tw2}{\small{$T_{\cW}^2$}}
\psfrag{tin}{\small{$\phi_0^d$}}
\psfrag{offset}{\small{offset}}
\includegraphics[width=6cm]{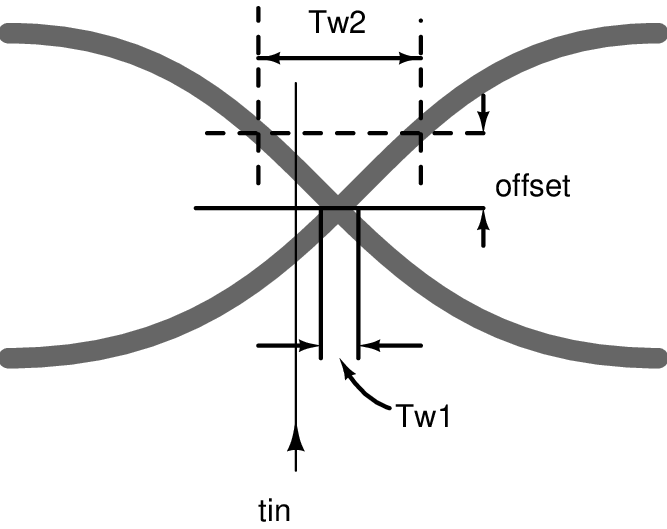}
\caption{Illustration of the sampler offset increasing the size of
the window \gls{cW} from $T_{\cW}^1$ to $T_{\cW}^2$.}
\label{fig:ch6:eye-withoffset}
\end{figure}

Linear phase detectors like the Hogge phase detector \cite{hogge}
behave similar to binary phase detectors in the window of susceptibility.
The ideal characteristics of the Alexander and Hogge phase detectors 
are shown in Fig.~\ref{fig:ch6:pd-chars}. The highlighted region shows 
the window \gls{cW}, and within this window the variation in the gain 
of the phase detector is negligible. Hence, the analysis of the system 
behaviour in the window \gls{cW} is applicable to both these types of 
phase detectors.
\textcolor{blue}{
The width of the window (\gls{TcW}) depends on the amount 
of timing jitter present in the system and the threshold of the 
sampling comparators. The effect of offset on \gls{TcW} is illustrated 
in Fig.~\ref{fig:ch6:eye-withoffset}.}
\textcolor{blue}{
To understand the effect of the offset, let us consider the case when 
there is an offset and the initial clock position at $\phi_0^d$ as 
shown in Fig.~\ref{fig:ch6:eye-withoffset}. %\subref{something}. 
Here we consider the scenario when there is non-zero offset, and 
the initial clock position $\phi_0^d$ is outside the window of 
susceptibility $\cW$ as shown in Fig.~\ref{fig:ch6:eye-withoffset}.  
%Note that the clock is actually outside the window of susceptbility 
%$\cW$ (according to our definition of $\cW$). 
Here, for 
every $0\rightarrow 1$ transition, the circuit makes a correction by 
shifting the clock to the left. Similarly, for every $1\rightarrow 0$ 
transition, the circuit makes a correction by shifting the clock to the 
right. Since every $0\rightarrow 1$ transition is followed by a 
$1\rightarrow 0$ transition, the net result is that the clock remains 
stuck. Hence, as shown in Fig.~\ref{fig:ch6:eye-withoffset}, offsets 
can result in increasing the width of the 
window of susceptibility $\cW$. We will revisit this with the example 
of offset in the presence of 1-bit ISI in 
Section~\ref{sec:offset_ISI}}.

\section{Experimental demonstration of the increase in settling time}
\label{sec:expt}
\begin{figure}[h!]
\centering
\psfrag{Vc}{\small{$V_c$}}
\psfrag{VH__}{\small{\hspace{-1ex}$V_H$}}
\psfrag{VL__}{\small{$V_L$}}
\psfrag{Timemicros}{\small{Time ($\mu s$)}}
\psfrag{Voltage}{\small{Voltage (V)}}
\psfrag{Phase}{\small{Phase}}
\psfrag{Detector}{\small{Detector}}
\psfrag{Weak}{\small{Weak}}
\psfrag{Strong}{\small{Strong}}
\psfrag{Charge}{\small{Charge}}
\psfrag{Pump}{\small{Pump}}
\psfrag{DLL}{\small{DLL}}
\psfrag{VCDL}{\small{VCDL}}
\psfrag{Fine}{\small{Fine tuning loop}}
\psfrag{Coarse}{\small{Coarse tuning loop}}
\psfrag{CkRx}{\small{$\phi_{Rx}$}}
\psfrag{Ckd}{\small{$\phi_d$}}
\psfrag{Data}{\small{Data}}
\psfrag{Input}{\small{Input}}
\psfrag{Control}{\small{Control}}
\includegraphics[width=0.45\columnwidth]{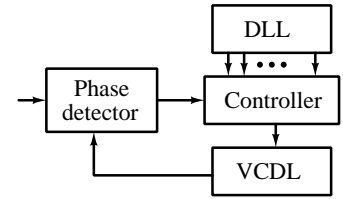}
\caption{Block diagram of the clock retiming circuit proposed in 
\cite{naveen_vlsi17}.}
\label{fig:cdr-coarse-fine}
\end{figure}

We use the coarse+fine retiming circuit for on-chip interconnects 
reported in~\cite{naveen_vlsi17} for the analysis and experiments. 
Fig.~\ref{fig:cdr-coarse-fine} 
shows the block diagram  of this synchronizer.

In this circuit, a delay locked loop (DLL) is used to generate multiple 
phases of the clock. A controller picks one of the phases of the DLL 
and delays it to bring the output clock to the center of the eye using 
a voltage controlled delay line (VCDL). The controller comprises a 
phase detector which senses the phase difference between the clock and 
the data and integrates the error using a charge pump. This charge pump 
is called the weak charge pump. The integrated error voltage is used 
to control the delay of the VCDL. If the VCDL range is not sufficient 
to achieve lock (which is detected by the control voltage $V_c$ 
exceeding preset upper and lower voltage bounds $V_H$ and $V_L$), the 
controller automatically picks the next adjacent phase, resets the 
control voltage using a strong charge pump and re-attempts to lock. 
The process repeats till lock is achieved.

\textcolor{blue}{
Most of the circuit blocks used in the clock retiming circuit 
in~Fig.~\ref{fig:cdr-coarse-fine} are standard circuit elements, 
except for the charge pump circuit which consists of the weak and 
the strong charge pumps. Fig.~\ref{fig:charge_pump} shows the 
transistor level circuit diagram of the weak and strong charge 
pumps.
\begin{figure}[h!]
\psfrag{Weak charge pump}{\small{Weak charge pump}}
\psfrag{Strong charge pump}{\small{Strong charge pump}}
\psfrag{UP}{\small{UP}}
\psfrag{DN}{\small{DN}}
\psfrag{DNB}{\small{$\overline{\text{DN}}$}}
\psfrag{UPB}{\small{$\overline{\text{UP}}$}}
\psfrag{1u}{\small{$1\mu A$}}
\psfrag{M1}{\small{$M_1$}}
\psfrag{M2}{\small{$M_2$}}
\psfrag{M3}{\small{$M_3$}}
\psfrag{M4}{\small{$M_4$}}
\psfrag{M5}{\small{$M_5$}}
\psfrag{M6}{\small{$M_6$}}
\psfrag{M7}{\small{$M_7$}}
\psfrag{M8}{\small{$M_8$}}
\psfrag{M9}{\small{$M_9$}}
\psfrag{M10}{\small{$M_{10}$}}
\psfrag{M11}{\small{$M_{11}$}}
\psfrag{M12}{\small{$M_{12}$}}
\psfrag{M13}{\small{$M_{13}$}}
\psfrag{M14}{\small{$M_{14}$}}
\psfrag{M15}{\small{$M_{15}$}}
\psfrag{UPS}{\small{UP$_{\text{st}}$}}
\psfrag{DNS}{\small{DN$_{\text{st}}$}}
\psfrag{Vc}{\small{$V_\text{c}$}}
\psfrag{vbn}{\small{$V_{bn}$}}
\psfrag{vbp}{\small{$V_{bp}$}}
\centering
\includegraphics[width=9cm]{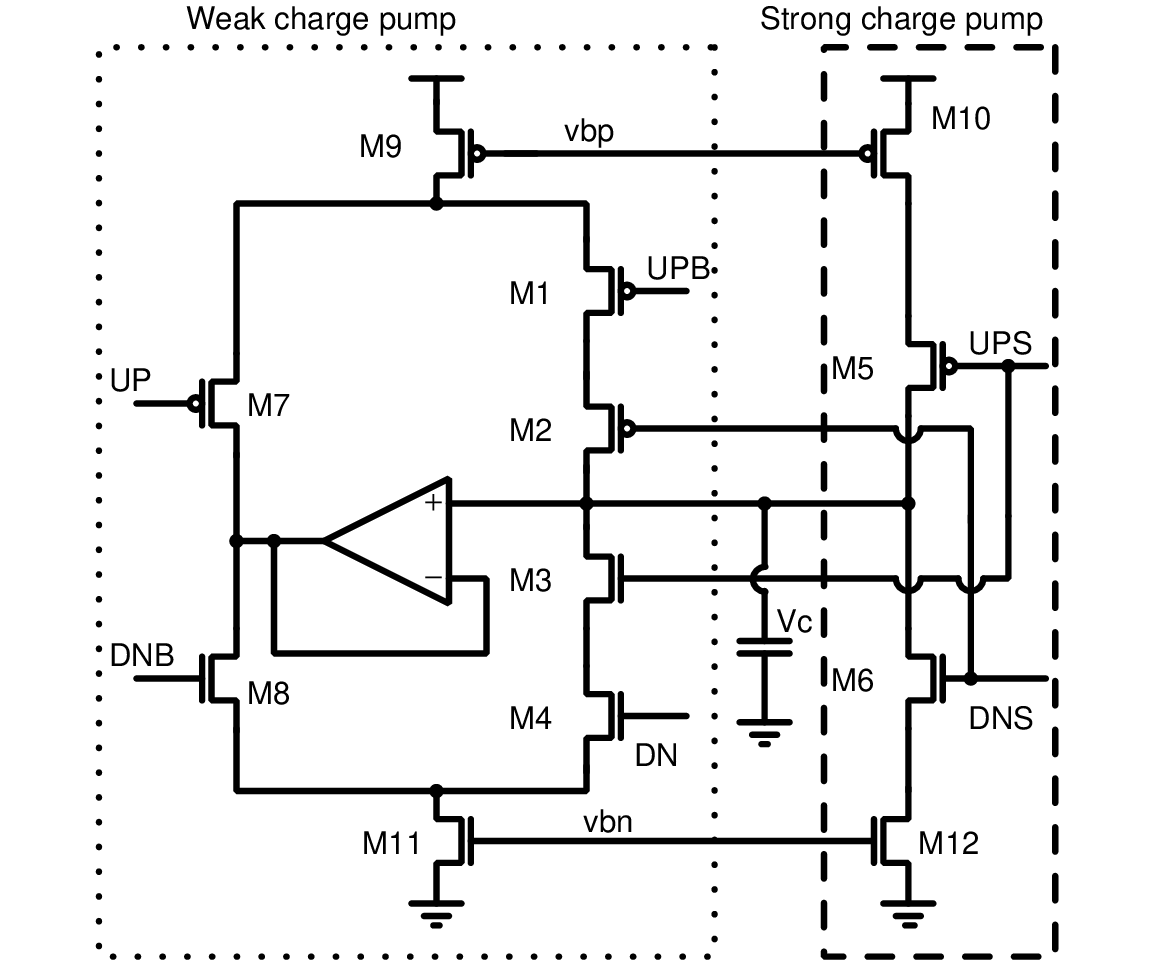}
\caption{\textcolor{blue}{Circuit diagram of the charge pump, which is 
made up of the the weak and the strong charge pumps. DN, $\overline{\text{DN}}$,
UP and $\overline{\text{UP}}$ signals come from the fine tuning loop and 
UP$_{\text{st}}$ and DN$_{\text{st}}$ signals come from the coarse 
tuning loop.}}
\label{fig:charge_pump}
\end{figure}
When the control voltage is outside the bounds $V_H$ and $V_L$, the 
strong charge pump is activated which brings the control voltage within 
the bounds. During this period, when the strong charge pump is active, 
the weak charge pump is disabled by the transistors $M_2$ and $M_3$ 
as shown in  Fig.~\ref{fig:charge_pump}.}

\begin{figure}[h!]
\centering
\psfrag{Tw}{\scriptsize{$T_{\cW}$}}
\psfrag{W}{\hspace{-0.3ex}\scriptsize{$\cW$}}
\psfrag{time (ns)}{\small{Two UI}}
\psfrag{receiver inp}{\small{\hspace{-2ex}Receiver i/p (mV)}}
\psfrag{Time (mus)}{\footnotesize{Time in $\mu$s}}
\psfrag{Voltage}{\footnotesize{Voltage}}
\psfrag{VH__}{\footnotesize{\hspace{1.5ex}$V_H$}}
\psfrag{Vc__}{\footnotesize{$V_c$}}
\psfrag{VL__}{\footnotesize{$V_L$}}
\psfrag{Clock stuck in W}{\footnotesize{Clock stuck in $\cW$}}
\includegraphics[width=8.5cm]{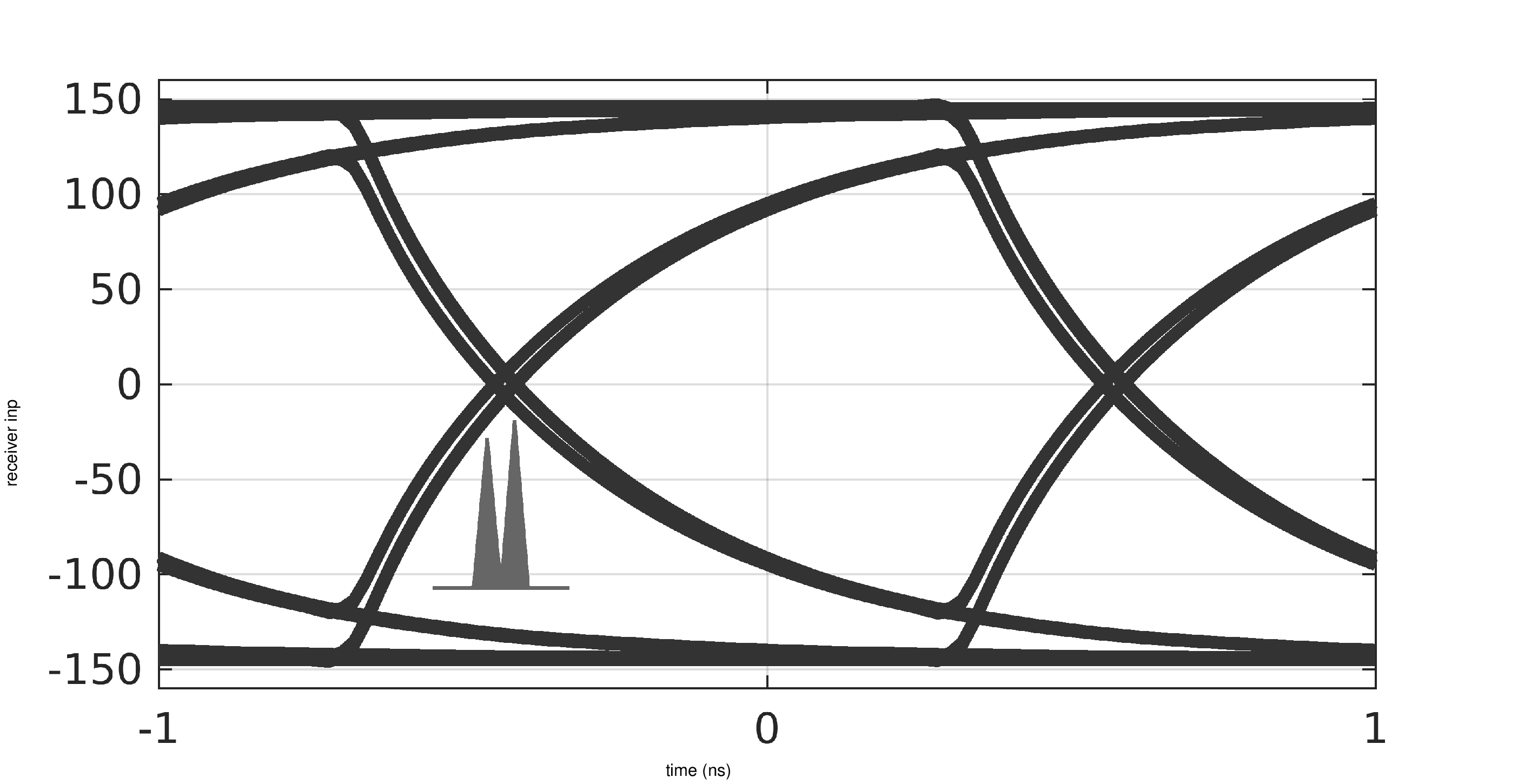}
\caption{Eye diagram at the receiver input along with the histogram of 
the zero crossing positions.}
%%of susceptibility.}
\label{fig:eye}
\end{figure}

We have measured the lengthening of the settling time on a prototype 
chip of this synchronizer. The chip has a circuit with a low swing 
transmitter and an interconnect which produces some ISI. 
Fig.~\ref{fig:eye} shows the simulated eye diagram at the receiver
for a clock frequency of 2~GHz.
%input of a typical low swing interconnect.
%Here, $T_{\cW}$ is the width of the window $\cW$. 
%

\begin{figure}[h!]
\centering
\psfrag{ck stuck}{\small{Clock stuck in $\cW$}}
\includegraphics[width=7.5cm]{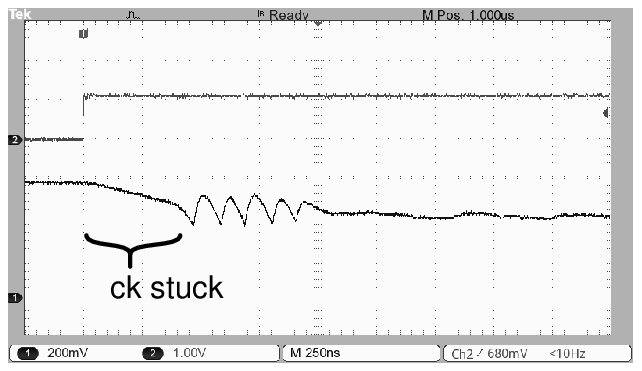}
\caption{Measured lengthening (by about 725 cycles) of the settling 
time, with the clock initialized within the window $\cW$.
The data is from a PRBS31 source with a Mark-to-Space 
ratio of 0.5. X-scale: 250~ns/div. Upper waveform is the reset 
(active low) signal.}
\label{fig:measured_extn}
\end{figure}

\begin{figure}[h!]
\centering
\includegraphics[width=7.5cm]{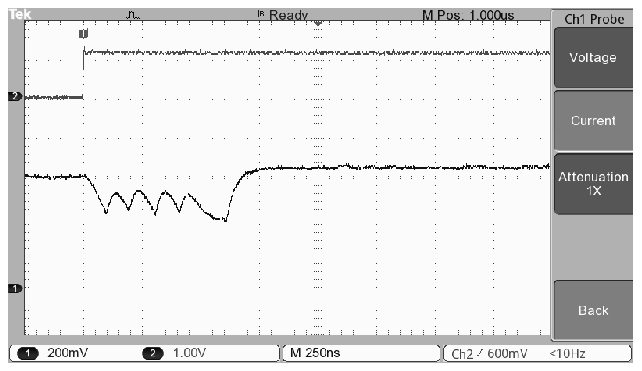}
\caption{Measured settling behaviour with the clock initialized at the 
edge of the window $\cW$. The data is from a PRBS31 source with a Mark-to-Space
ratio of 0.5. X-scale: 250~ns/div. Upper waveform is the reset 
(active low) signal.}
\label{fig:measured_noextn}
\end{figure}

The chip was tested at a frequency of 2~GHz with Centellax TG1B1-A data 
generator~\cite{ber-tester} and Centellax TG1C1-A clock 
synthesizer~\cite{ck-generator}. In the clock synthesizer instrument, 
one of the clock outputs can be phase shifted with a resolution 
$2^{\circ}$, which is used for programming the initial phase error. 
Fig.~\ref{fig:measured_extn} shows the control voltage evolution when 
the circuit is stuck in the horizontally closed region of the eye at 
startup.
Notice the abrupt change in the slope of the signal when the circuit 
escapes the window $\cW$, which happens after about 725 cycles in this 
example. When the circuit is initialized at the edge of this window, 
such a lengthening of settling time is not observed as shown in 
Fig.~\ref{fig:measured_noextn}.

\begin{figure}[h!]
\centering
\psfrag{ck stuck}{\small{Clock stuck in $\cW$}}
\includegraphics[width=7.5cm]{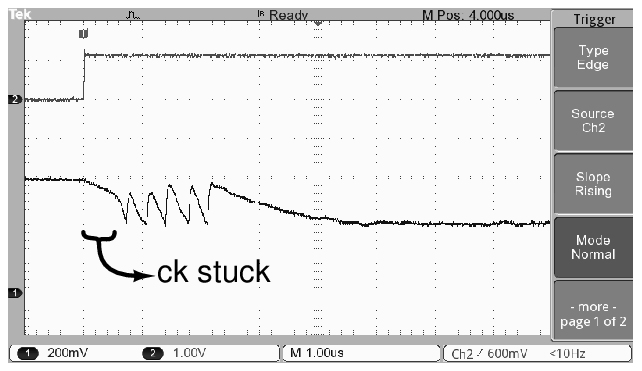}
\caption{Measured lengthening of the settling time (about 1200 cycles), 
with the clock initialized within the window $\cW$. The data is from 
a PRBS31 source with a Mark-to-Space ratio of 0.125. 
X-scale: 1~$\mu$s/div. Upper waveform is the reset 
(active low) signal.}
\label{fig:measured_extn_MS125}
\end{figure}

The lengthening of the settling time was observed with different data 
patterns like PRBS7, PRBS15 and PRBS31. In addition, the TG1B1-A data 
generator produces outputs with different Mark-to-Space ratios. The 
Mark-to-Space ratios can be chosen between $0.5$, $0.25$ and $0.125$. 
Even with these different patterns, the increase in settling time 
occurs when the clock is initialized in the closed region of the data 
eye. Fig.~\ref{fig:measured_extn_MS125} shows the control voltage 
evolution when the circuit is stuck in the horizontally closed region 
of the eye with a data pattern with Mark-to-Space ratio of $0.125$, in 
which case the circuit remains stuck with its clock in the window $\cW$ 
for about 1200 cycles.
As expected the total settling time is higher for a lower data activity 
factor. It is worth noting that changing the Mark-to-Space ratio only 
changes the data activity factor. The minimum run length for logic 1 
and logic 0 is still 1-bit. We will show in Section~\ref{sec:reduce} 
that by changing the 
minimum run length for one of the logic levels, the settling time can 
be reduced considerably.

Having demonstrated the problem of increased settling time, the next 
section will analyze the settling time of these circuits in the 
presence of different types of jitter.

%% file: markov.tex
\section{Markov chain model of the clock recovery circuit}
\label{sec:markov}
In order to analyze the settling time when the initial clock is in the 
window \gls{cW}, the circuit is modelled as a Markov chain with 
absorbing states~\cite{markov}. 
The binary phase detector produces UP or DN pulses on every 
data transition, which are converted into an analog control voltage 
using a charge pump. The control voltage is used to delay the clock 
(either using a phase interpolator or a delay line). Since the input 
is binary, the output phase is quantized. Hence, the clock position 
can be discretized to the step size ($\tau$) of the phase detector loop 
update. In order to keep the jitter in the recovered clock low, 
controllers typically use step sizes of less than 0.1\% of the clock 
period 
(c.f.,~\cite{naveen_vlsi17,phase_interpolator,phase_resn1,phase_resn2}).
The region where the input data eye is closed, i.e., the window $\cW,$
is of particular interest. A Markov chain model of the system is 
constructed in which the states designate the clock positions and the 
phase corrections performed by the clock retiming circuit form its 
state transitions. The edges of the window \gls{cW} are modelled as 
absorbing states.

The sources of timing jitter are data dependent jitter (\gls{isi} 
induced) in the data and noise induced random jitter in the clocks of 
the transmitter and receiver. The analysis for each type of jitter is 
done separately, followed by an analysis for data dependent and random 
jitter together. For analyzing the effect of data dependent jitter, an 
interconnect link with different bandwidths is considered. The 
interconnect is modelled as a 20 section RC network. For simulating 
different amounts of ISI, different values of RC time constants are 
chosen and the eye diagrams and timing jitter histograms for a few
considered channel bandwidths are listed as cases 1 through 3 as 
follows: \\ 

\begin{description}
\item[Case 1\hspace{1ex}] For benign channels the horizontal eye opening
approaches 100\% and the jitter histogram has a narrow distribution as 
shown in Fig.~\ref{fig:ch6:eye-benign}. 
\begin{figure}[h!]
\centering
\psfrag{data}{\small{Receiver i/p (mV)}}
\psfrag{Time}{\small{One UI}}
\includegraphics[width=8.5cm]{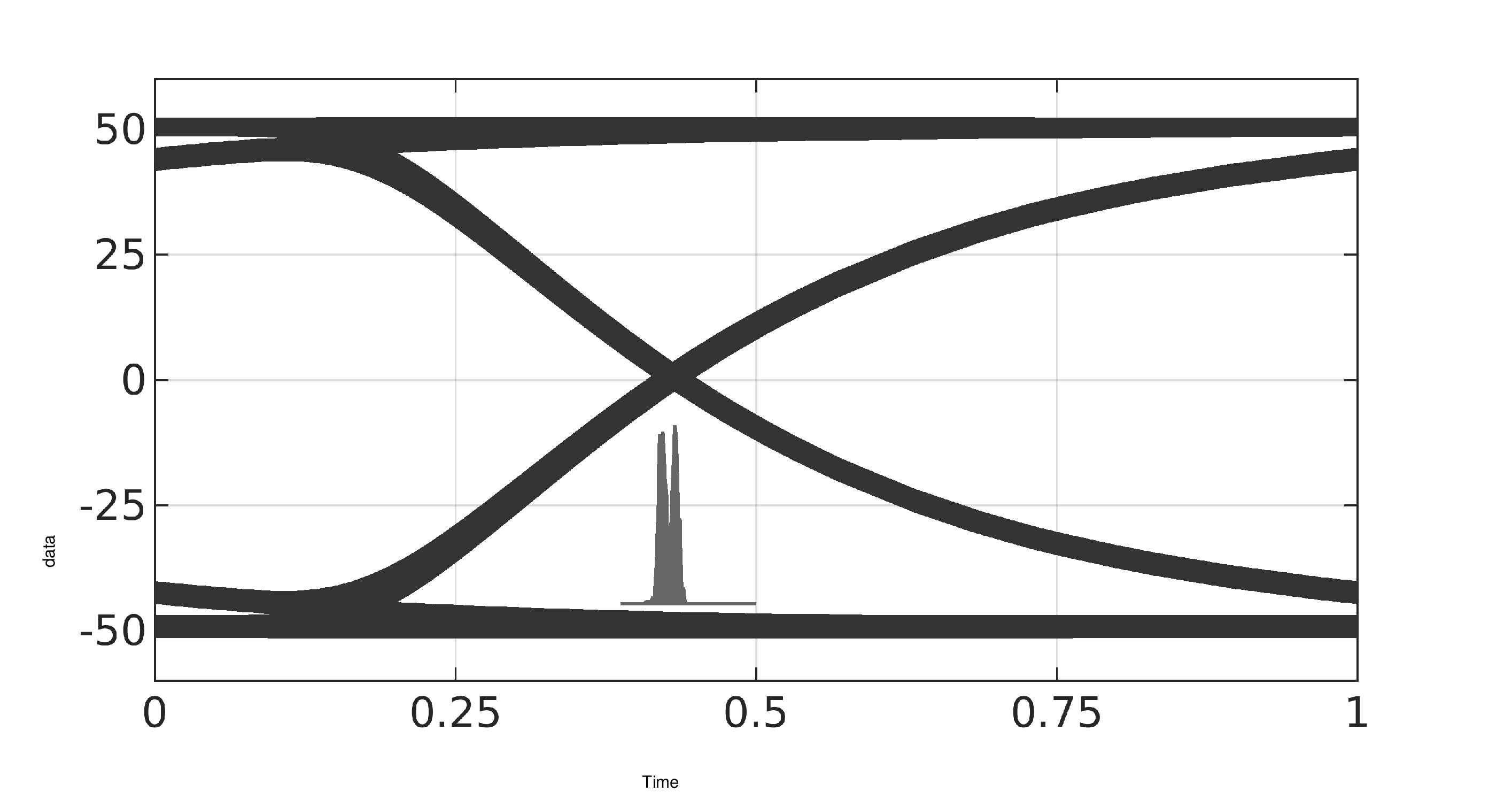}
\caption{Eye diagram and zero crossing histogram for a benign channel.}
\label{fig:ch6:eye-benign}
\end{figure}
\item[Case 2\hspace{1ex}] As the bandwidth of the channel decreases, the
jitter histogram splits and produces two distinct 
peaks~\cite{ddj_prediction}. This shows that the ISI due to the 
immediate previous bit is dominant (Fig.~\ref{fig:ch6:moderate_isi}). 
\begin{figure}[h!]
\centering
\psfrag{data}{\small{Receiver i/p (mV)}}
\psfrag{Time}{\small{One UI}}
\psfrag{A}{\small{$\cA$}}
\psfrag{B}{\small{$\cB$}}
\includegraphics[width=8.5cm]{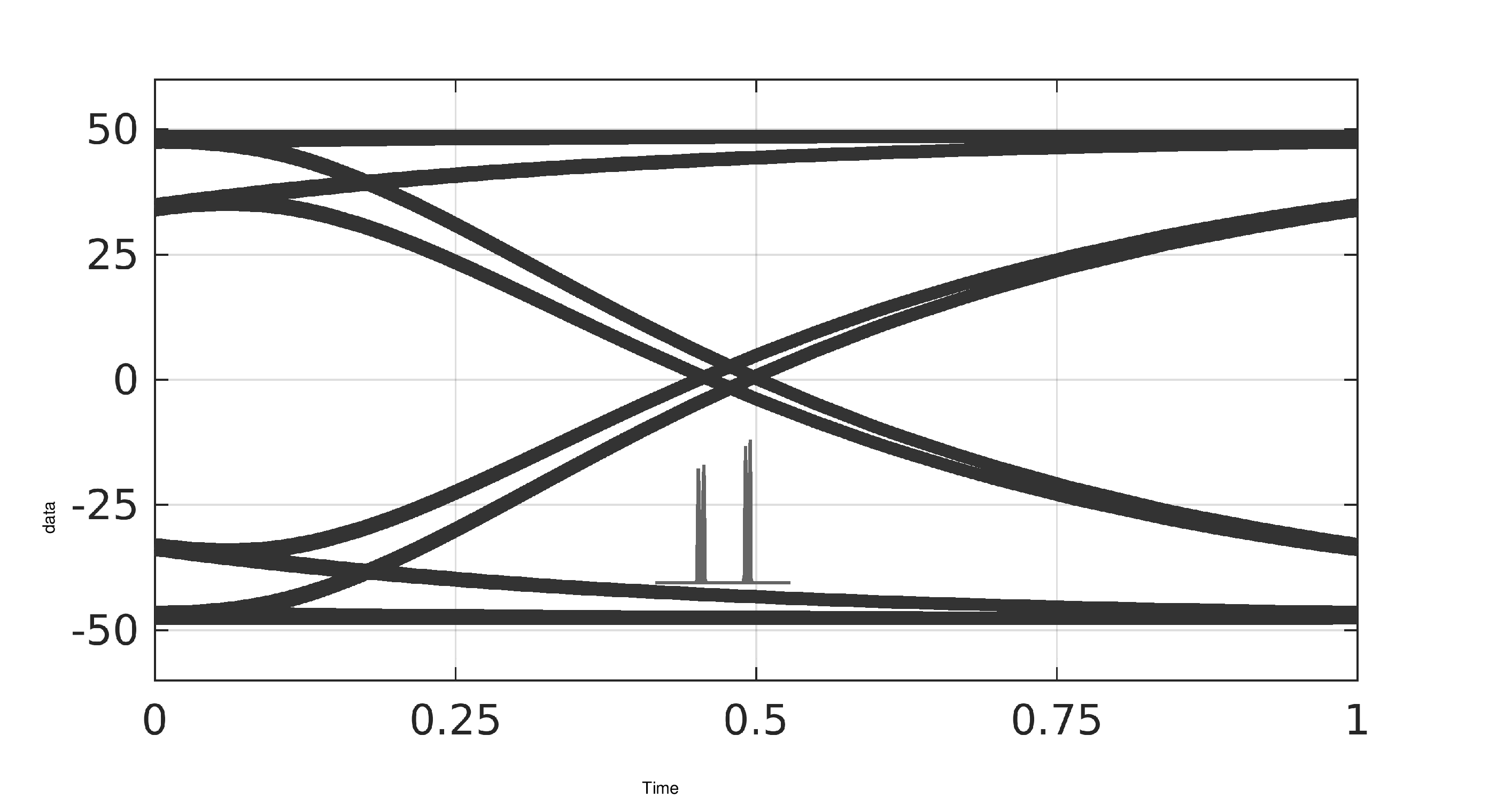}
\caption{Eye diagram and zero crossing histogram for a channel for which the
ISI due to the immediate previous bit is dominant.}
\label{fig:ch6:moderate_isi}
\end{figure}
\item[Case 3\hspace{1ex}] Further reduction in the  bandwidth shows that 
the jitter histogram splits into 4 regions as shown in 
Fig.~\ref{fig:ch6:high_isi}. This shows that the ISI due to previous 
two bits is dominant. 
\begin{figure}[h!]
\centering
\psfrag{data}{\small{Receiver i/p (mV)}}
\psfrag{Time}{\small{One UI}}
\psfrag{A}{\small{$\cA$}}
\psfrag{B}{\small{$\cB$}}
\psfrag{C}{\small{$\cC$}}
\psfrag{D}{\small{$\cD$}}
\includegraphics[width=8.5cm]{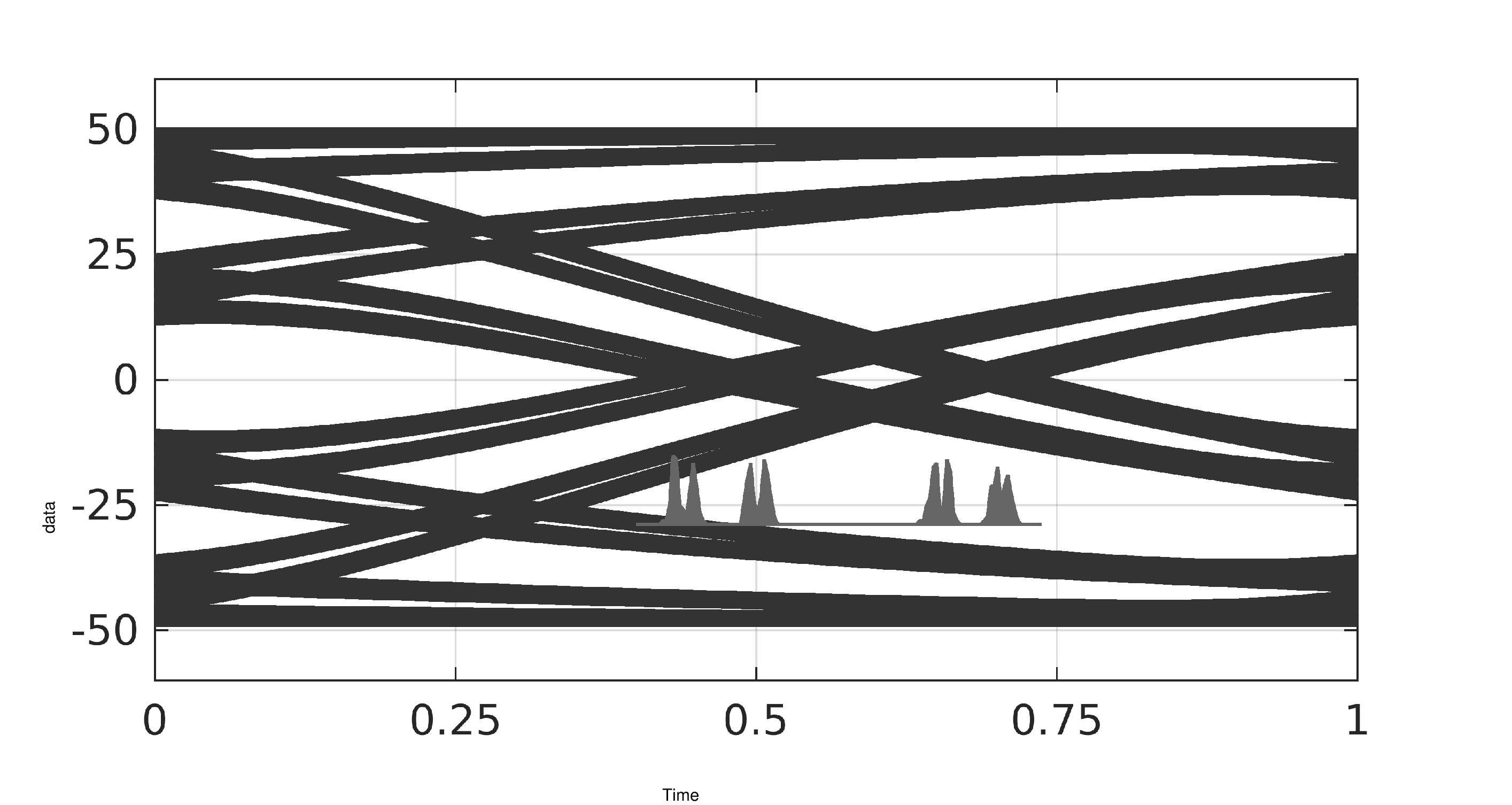}
\caption{Eye diagram and zero crossing histogram for a channel for 
which the ISI due to previous 2 bits is dominant.}
\label{fig:ch6:high_isi}
\end{figure}
\end{description}

Further reduction in the bandwidth of the channel results in the 
jitter histogram splitting to 8 peaks, 16 peaks and so 
on~\cite{ddj_prediction}. While the modeling presented in this paper 
can be used to analyze these scenarios, such channels are not usable 
due to the very high ISI present. Hence, our analysis considers ISI 
of 1 previous bit and 2 previous bits. For modeling effect of random 
jitter, the random jitter is assumed to have a Gaussian 
distribution~\cite{generated_jitter}. Overall, the system is 
analyzed for the cases when the jitter is 
\begin{enumerate}%[\it A.]
\item induced by ISI due to 1 previous bit,
\item induced by ISI due to 2 previous bits,
\item random with a Gaussian distribution and 
\item induced by random jitter and \gls{isi} due to 1 previous bit.
\end{enumerate}

\subsection{Markov chain model for jitter induced by 1 bit ISI}
\label{subsec:ch6:markovmodel-1bit}
\begin{figure}[h!]
\centering
\psfrag{Data}{\small{Data}}
\psfrag{Clock}{\small{Clock}}
\psfrag{A}{\small{$\cA$}}
\psfrag{B}{\small{$\cB$}}
\psfrag{ti}{\small{$t_i$}}
\psfrag{t=0}{\small{$t=0$}}
\psfrag{2Tw}{\small{$T_{\cW}$}}
\psfrag{tin}{\small{$t_i$}}
\psfrag{Vth}{\small{$V_{th}$}}
\psfrag{alph}{\small{$\gamma$}}
\includegraphics[width=4cm]{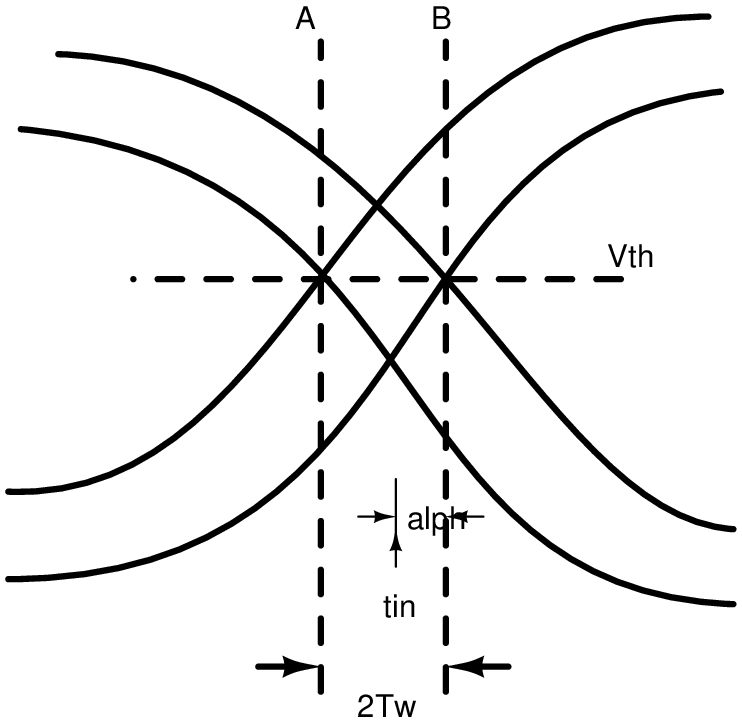}
\caption{Sketch of an eye diagram with ISI limited to 1 previous 
bit.}
\label{fig:ch6:eye_cartoon}
\end{figure}

When the \gls{isi} due to the immediate previous bit is dominant, the 
zero crossings of the data are bunched into two narrow distributions 
as shown in Fig.~\ref{fig:ch6:moderate_isi}. This is approximated to 
the data signal following one of two distinct traces. 
Fig.~\ref{fig:ch6:eye_cartoon} illustrates an eye diagram with 
\gls{isi} due to exactly 1 previous bit.
Here $t_i$ is the initial sampling instant of the clock, $\gamma$ is 
the distance to the right edge of the window and $V_{th}$ is the
threshold of the samplers in the phase detector. $\cA$ and $\cB$ 
represent the two distinct zero crossing times of the data signal. For 
the eye diagram shown in Fig.~\ref{fig:ch6:moderate_isi}, the size of 
the window $\cW$ is about 4\% of the bit period. This means $T_{\cW}$ 
is about $40\tau$. 
Assuming that the source outputs 1 and 0 with equal probability, the 
bit combinations that result in traces with zero crossing at $\cA$ and
$\cB$, respectively, are listed in Table~\ref{tbl:ch6:isi_1bit}. Here 
$b_0$ is the current bit corrupt with ISI due to immediately previous 
bit $b_{-1}$. A data transition occurs when $b_0 \neq b_{1}$. 
Table~\ref{tbl:ch6:isi_1bit} lists all 3 bit combinations which cover 
all possible data traces. 
\begin{table}[h]
\begin{center}
\small{
\caption{Possible sequences and data traces for 1 bit ISI.}
\label{tbl:ch6:isi_1bit}
\begin{tabular}{|ccc|c|c|}
\hline
$b_{-1}$ & $b_0$ & $b_{1}$ & Zero crossing time & Action \\ \hline \hline
0 & 0 & 0 & - & $NA$ \\ \hline
0 & 0 & 1 & $\cB$ &  $LT$ \\ \hline
0 & 1 & 0 & $\cA$ & $RT$ \\ \hline
0 & 1 & 1 & - & $NA$ \\ \hline
1 & 0 & 0 & - & $NA$ \\ \hline
1 & 0 & 1 & $\cA$ & $RT$ \\ \hline
1 & 1 & 0 & $\cB$ & $LT$ \\ \hline
1 & 1 & 1 & - & $NA$ \\ \hline
\end{tabular}}
\end{center}
\end{table}

Here $LT$ and $RT$ indicate that the clock is shifted to the left
and to the right, respectively, by a step of size $\tau$ while $NA$ 
indicates no corrective action in that cycle. When all sequences are 
equally likely, the probabilities $\prob(RT) = 0.25 = \prob(LT)$  and 
$\prob(NA) = 0.5$. Note that the system escapes the window of 
susceptibility $\cW$ when, for the first time, either 
\mbox{$(n_R - n_L)\tau \ge \gamma$} or 
\mbox{$(n_L - n_R)\tau \ge (T_{\cW} - \gamma)$}. 
Here $n_R$ and $n_L$ are the total number of times the clock position 
has been shifted to the right and to the left, respectively, 
from start-up.

\begin{figure}[h!]
\centering
\psfrag{p0}{\small{$\frac{1}{2}$}}
\psfrag{p+}{\small{$\frac{1}{4}$}}
\psfrag{p-}{\small{$\frac{1}{4}$}}
\psfrag{k}{\small{$T_{\cW}/2$}}
\psfrag{-k}{\small{$-T_{\cW}/2$}}
\psfrag{1}{\small{1}}
\psfrag{0}{\small{0}}
\includegraphics[width=0.475\columnwidth]{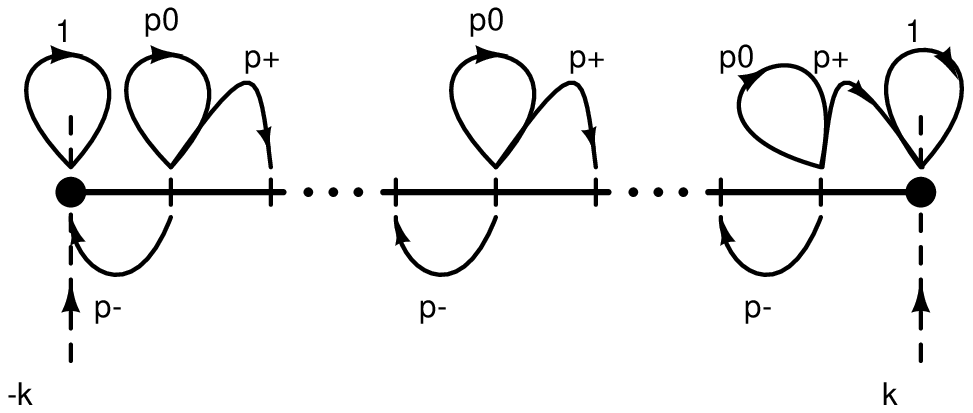}
\caption{Model of the clock recovery circuit, when the data has 
ISI limited to 1 previous bit, as a Markov chain with 
absorbing states.} 
\label{fig:ch6:mc_model}
\end{figure}

This system is modelled as a one dimensional Markov chain. The states of 
the Markov chain correspond to the positions of the sampling clock
in the window $\cW.$ Once the clock escapes this window, the time 
taken to lock to the center of the eye can be calculated 
from~\eqref{eqn:ch6:ts_deterministic}. 
Thus, the edge positions of the window are modelled as \emph{absorbing} 
states, thereby making this a Markov chain with absorbing 
states~\cite{markov}. Fig.~\ref{fig:ch6:mc_model} 
shows the state diagram of the Markov chain for this system. 
By knowing the state space and the transition probabilities of a Markov 
chain, one can calculate the mean time to absorption from any initial 
state~\cite{markov}. The calculation of the mean time to absorption 
(and its variance) is presented in Appendix~\ref{apdx:markov-meantime}. 

\begin{figure}[h!]
\centering
\psfrag{Behavioural}{\scriptsize{Behavioural}}
\psfrag{simulation}{\scriptsize{Simulation}}
\psfrag{Model}{\scriptsize{Model}}
\psfrag{prediction}{\scriptsize{prediction}}
\psfrag{Initial clock position}{\small{Initial clock position}}
\psfrag{Absorbtion Time (#cycles)}{\small{\hspace{-2ex} 
Mean absorption time (cycles)}}
\psfrag{Tw}{\scriptsize{$T_{\cW}/2$}}
\psfrag{-Tw}{\scriptsize{-$T_{\cW}/2$}}
%\subfloat[Mean absorption time]{
%\includegraphics[width=8.2cm]{../figs/mean_abs_beh_model}
%\label{fig:ch6:markovmodel-1bit-mean-abs}}
%\hspace{1ex}
\psfrag{Behavioural}{\scriptsize{Behavioural}}
\psfrag{simulation}{\scriptsize{Simulation}}
\psfrag{Model}{\scriptsize{Model}}
\psfrag{prediction}{\scriptsize{prediction}}
\psfrag{Initial clock position}{\small{Initial clock position}}
\psfrag{standard deviation (#cycles)}{\small{Standard deviation (cycles)}}
\psfrag{Tw}{\scriptsize{$T_{\cW}/2$}}
\psfrag{-Tw}{\scriptsize{-$T_{\cW}/2$}}
%\begin{subfigure}[b]{0.49\textwidth}
%\centering
\includegraphics[width=0.45\columnwidth]{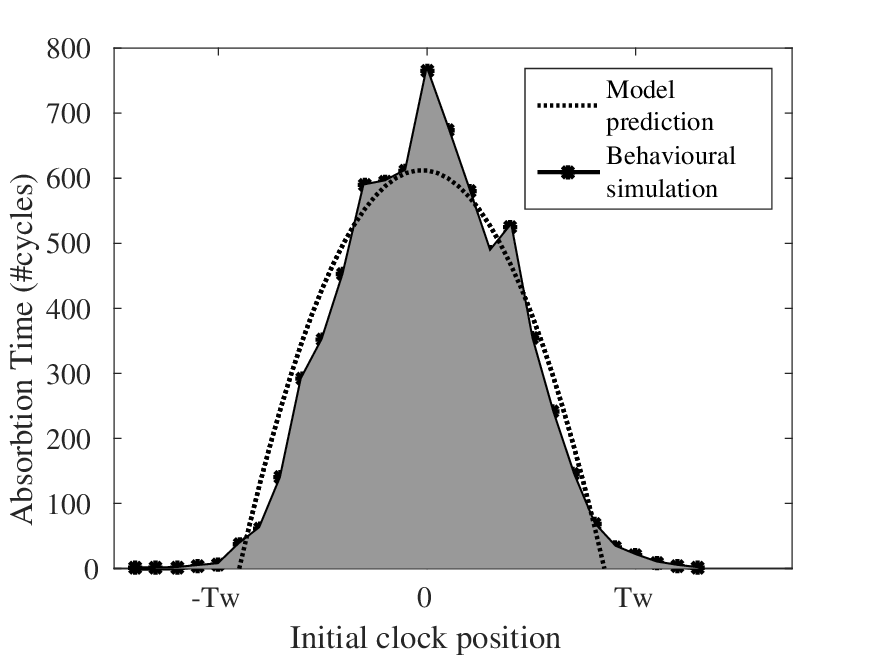}
%\subcaption{Mean absorption time}
%\end{subfigure}
%\begin{subfigure}[b]{0.49\textwidth}
%\centering
%\subfloat[Standard deviation of absorption time]{
%\includegraphics[width=0.9\textwidth]{../figs/std_abs_beh_model}
%\subcaption{Standard deviation of absorption time}
%\label{fig:ch6:markovmodel-1bit-std-abs}
%\end{subfigure}
\caption{Mean absorption time as 
predicted from the Markov chain model and from behavioural
simulations. Each data point in the behavioural simulation is an average
of 100 runs.}
\label{fig:ch6:markovmodel-1bit-mean-abs}
%\label{fig:ch6:markovmodel-1bit}
\end{figure}

The combined plots of the mean settling time for the 1 bit ISI case 
obtained from behavioural simulations of the circuit and its Markov 
chain model predictions via the mean absorption time are shown in 
Fig.~\ref{fig:ch6:markovmodel-1bit-mean-abs}.
The behavioural simulations were done using a 20 section RC interconnect
and a VerilogA behavioural description of the clock retiming circuit, 
while the model predictions were computed by solving for the mean 
absorption time of this Markov chain using a linear equation solver.
%while the 
%behavioural simulations were done using a 20 section RC interconnect 
%and a VerilogA behavioural description of the clock retiming circuit. 

\begin{figure}[h!]
\centering
\psfrag{Behavioural}{\scriptsize{Behavioural}}
\psfrag{simulation}{\scriptsize{Simulation}}
\psfrag{Model}{\scriptsize{Model}}
\psfrag{prediction}{\scriptsize{prediction}}
\psfrag{Initial clock position}{\small{Initial clock position}}
\psfrag{Tw}{\scriptsize{$T_{\cW}/2$}}
\psfrag{-Tw}{\scriptsize{-$T_{\cW}/2$}}
\psfrag{standard deviation (#cycles)}{\small{Standard deviation (cycles)}}
\includegraphics[width=0.45\columnwidth]{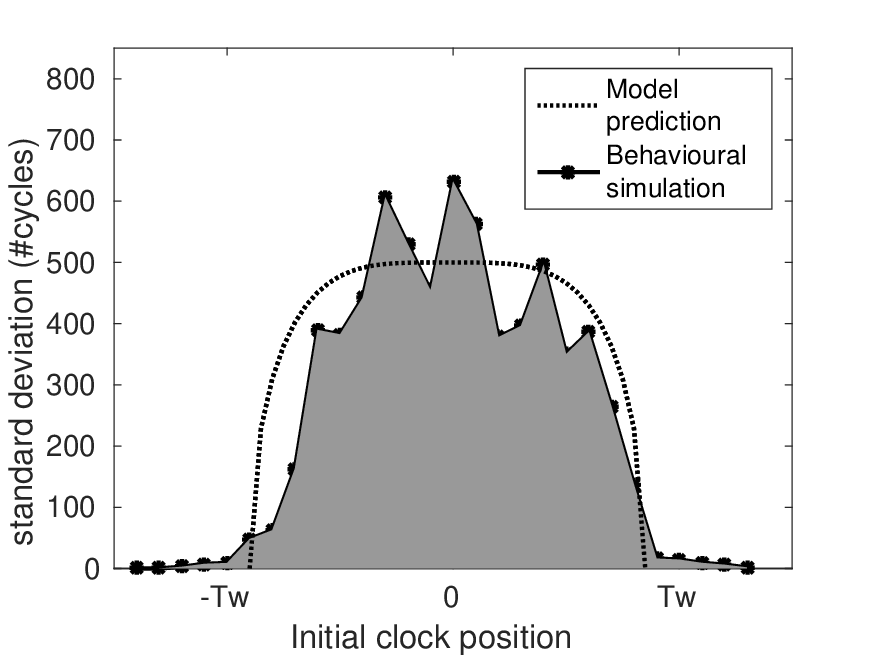}
\caption{Standard deviation of the absorption time as 
predicted from the Markov chain model and from behavioural
simulations. Each data point in the behavioural simulation is an average
of 100 runs.}
\label{fig:ch6:markovmodel-1bit-std-abs}
\end{figure}

As one would expect, the settling time is maximum when the initial 
sampling phase is at the center of the window $\cW.$ It is worth 
noting that the variance of the absorption time is quite high. 
Fig.~\ref{fig:ch6:markovmodel-1bit-std-abs} shows the standard 
deviation of the absorption time as predicted by the Markov chain model 
and as observed in the data obtained from behavioural simulations.

\subsubsection{Modeling the effect of offset and 1-bit ISI}
\label{sec:offset_ISI}
\begin{figure}[h!]
\centering
\psfrag{Data}{\small{Data}}
\psfrag{Clock}{\small{Clock}}
\psfrag{A}{\small{$\cA$}}
\psfrag{A1}{\small{$\cA'$}}
\psfrag{B}{\small{$\cB$}}
\psfrag{B1}{\small{$\cB'$}}
\psfrag{ti}{\small{$t_i$}}
\psfrag{t=0}{\small{$t=0$}}
\psfrag{2Tw}{\small{$T_{\cW}$}}
\psfrag{tin}{\small{$t_i$}}
\psfrag{Vth}{\small{$V_{th}$}}
\psfrag{alph}{\small{$\gamma$}}
\includegraphics[width=4cm]{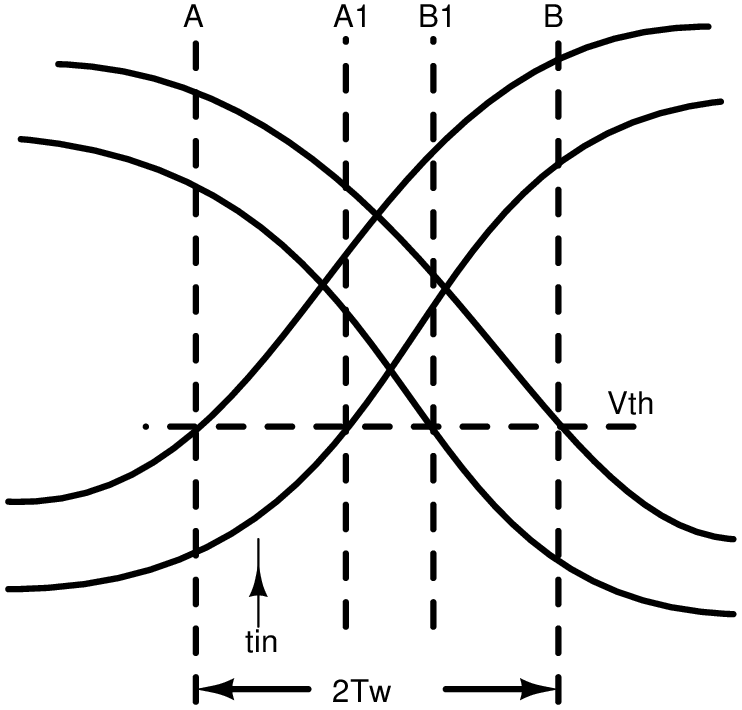}
\caption{\textcolor{blue}{Sketch of an eye diagram illustrating 
the effect of offset for the case when the ISI is limited to 
1 previous bit.}}
\label{fig:ch6:eye_cartoon_with_offset}
\end{figure}

\textcolor{blue}{
As was discussed in Section~\ref{sec:settling}, offsets can increase 
the width of the window of susceptibility. 
Fig.~\ref{fig:ch6:eye_cartoon_with_offset} illustrates an example of 
the effect of offset in the case when the ISI is limited to 1 
previous bit.}
\textcolor{blue}{
In this case, the circuit behaviour crucially depends on the 
clock position in the window. In particular, it differs when 
the clock is in $\cA - \cA'$, $\cA' - \cB'$ and in $\cB' - \cB$.  
%In this case, the circuit behaviour will be different from when the 
%clock is in the window from $\cA - \cA'$, $\cA' - \cB'$ and in 
%$\cB' - \cB$. 
For instance, using a procedure similar to the one 
described for analyzing 1-bit ISI, one can easily verify that 
the probabilities 
for left and right shifts are as follows.}

\begin{table}[h!]
\begin{center}
{\color{blue}
\begin{tabular}{|c|c|c|c|}
%\textcolor{blue}
\hline
Window & $\prob (LT)$ & $\prob (RT)$ & $\prob (NA)$ \\ \hline
$\cA - \cA'$ & 0.375 & 0.125 & 0.5 \\ \hline 
$\cA' - \cB'$ & 0.25 & 0.25 & 0.5 \\ \hline 
$\cB' - \cB$ & 0.125 & 0.375 & 0.5 \\ \hline 
\end{tabular}}
\end{center}
\end{table}

\textcolor{blue}{
By constructing the Markov chain with above transition probabilities, 
the settling time analysis can be extended to analyze 
the effect of any given offset in a straightforward manner. 
}

\subsection{Markov chain model for jitter induced by 2 bits ISI}
\begin{figure}[h!]
\centering
\psfrag{A}{\small{$\cA$}}
\psfrag{B}{\small{$\cB$}}
\psfrag{C}{\small{$\cC$}}
\psfrag{D}{\small{$\cD$}}
\psfrag{000}{\small{$000$}}
\psfrag{001}{\small{$001$}}
\psfrag{010}{\small{$010$}}
\psfrag{011}{\small{$011$}}
\psfrag{100}{\small{$100$}}
\psfrag{101}{\small{$101$}}
\psfrag{110}{\small{$110$}}
\psfrag{111}{\small{$111$}}
\includegraphics[width=7cm]{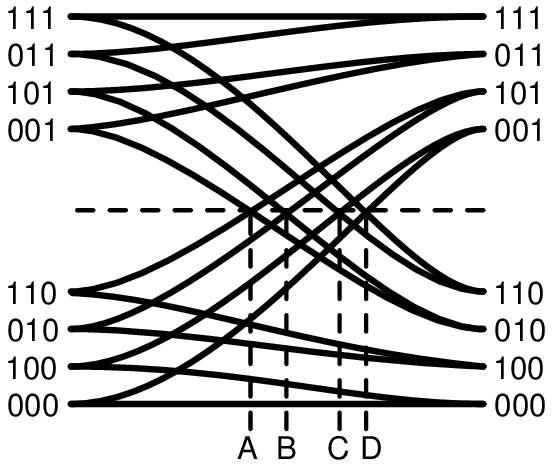}
\caption{Sketch of an eye diagram with ISI due to previous 2 bits, 
showing the four discrete data transition times.}
\label{fig:ch6:eye_cartoon_2bit}
\end{figure}

When the \gls{isi} due to previous two bits is significant, the data 
transitions histogram has 4 peaks as shown in 
Fig.~\ref{fig:ch6:high_isi}. This can be approximated to ISI of exactly 
two bits which results in 4 distinct data transition times. 
Fig.~\ref{fig:ch6:eye_cartoon_2bit} 
illustrates an eye diagram when ISI is limited to two bits and the
four data transition times are labeled as $\cA$, $\cB$, $\cC$ and $\cD$. 
Thus, the window of susceptibility $\cW$ can be divided in to three 
sub-windows which are $\cW_{\cA-\cB}$, $\cW_{\cB-\cC}$ and 
$\cW_{\cC-\cD}$. Unlike the 1 bit \gls{isi} case, the occurrences of 
the traces with zero crossing time at $\cA$, $\cB$, $\cC$ and $\cD$ are 
not independent of each other. Thus, the system is no longer 
memoryless, and hence, requires a second order Markov chain model. 
A second order Markov chain model is, in general, difficult to 
analyze. A technique to convert this second order Markov chain to a 
first order Markov chain (with an enlarged state space) is 
described in the following. This greatly simplifies the analysis, as 
known results for first order Markov chains can be directly applied.

\begin{figure}[h!]
\centering
	\begin{tikzpicture}[->,thick,>=stealth',node distance=2.3cm]
\tikzstyle{every state}=[align=center]
\tikzstyle{every node}=[font=\small]

\node[state](S0){$S_0$\\000};
\node[state](S1)[below right of=S0,yshift=0.5cm]{$S_1$\\001};
\node[state](S2)[right of=S1]{$S_2$\\010};
\node[state](S3)[below left of=S1,yshift=-0.5cm,xshift=-1cm]{$S_3$\\011};
\node[state](S4)[below of=S2,xshift=-1cm]{$S_4$\\100};
\node[state](S5)[below right of=S4]{$S_5$\\101};
\node[state](S6)[below left of=S4,yshift=-1cm]{$S_6$\\110};
\node[state](S7)[below left of=S6,xshift=-1cm,yshift=2cm]{$S_7$\\111};

\path(S0)edge[loop above]node       {\small{0/$\cX^{(2)}$}}(S0);
\path(S0)edge            node[above]{\small{1/$\cD$}}(S1);

\path(S1)edge            node[above]{\small{0/$\cA$}}(S2);
\path(S1)edge            node[left]{\small{1/$\cX^{(1)}$}}(S3);

\path(S2)edge            node[right]{\small{0/$\cX^{(1)}$}}(S4);
\path(S2)edge[bend left=58]node[right]{\small{1/$\cB$}}(S5);

\path(S3)edge            node[right]{\small{0/$\cC$}}(S6);
\path(S3)edge            node[left]{\small{1/$\cX^{(2)}$}}(S7);

\path(S4)edge[bend left=40]node[right]{\small{0/$\cX^{(2)}$}}(S0);
\path(S4)edge node[right]{\small{1/$\cC$}}(S1);

\path(S5)edge[bend right=20]node[right]{\small{0/$\cB$}}(S2);
\path(S5)edge            node[above]{\small{1/$\cX^{(1)}$}}(S3);

\path(S6)edge            node[right]{\small{0/$\cX^{(1)}$}}(S4);
\path(S6)edge            node[below]{\small{1/$\cA$}}(S5);

\path(S7)edge            node[below]{\small{0/$\cD$}}(S6);
\path(S7)edge[loop below]node{\small{1/$\cX^{(2)}$}}(S7);

\end{tikzpicture}

\caption{State diagram indicating possible transitions and data zero 
crossing positions, for data corrupted with ISI from previous 2 bits.}
%$\cX^{(i)}$,$i=1,2$ indicates no data transition 
%in the previous and previous 2 cycles resectively.}
\label{fig:ch6:twobit_fsm}
\end{figure}
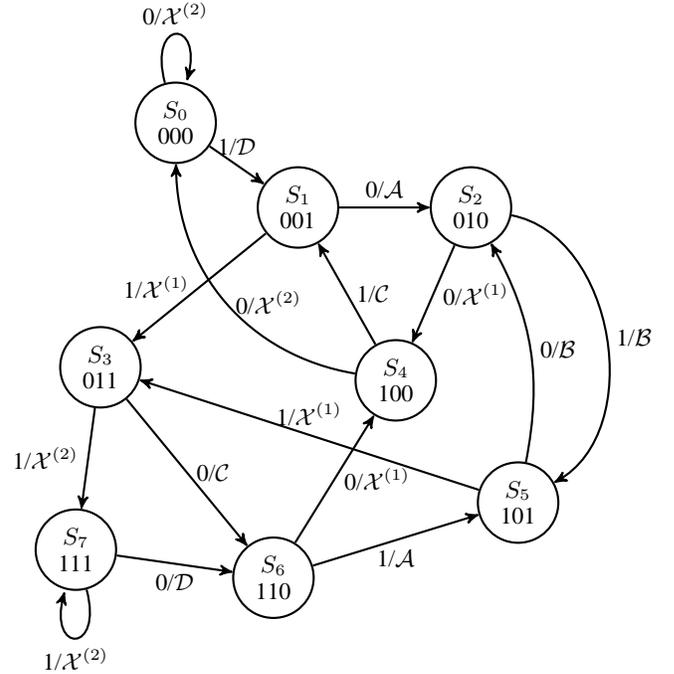

To construct the Markov chain model of the system, one needs to first 
consider the order of occurrences of the traces with zero crossing 
times at $\cA$, $\cB$, $\cC$ and $\cD$. Fig.~\ref{fig:ch6:twobit_fsm} 
shows the state diagram of the source data. 
The \gls{fsm} is limited to 3 bit 
combinations as the ISI for every bit is limited to previous two bits. 
The state transitions are labeled with the zero crossing locations of 
the data signal $\cA$ through $\cD$. $\cX^{(i)}$, $i=1,2$, indicates a 
state transition without a data transition.

\begin{figure}[h!]
\centering
\begin{tikzpicture}[->,>=stealth,thick,node distance = 1.25cm]
%[->,thick,>=stealth',node distance=1cm]
%\tikzstyle{every state}=[align=center,fill=none,draw=black,thick,text=black]
%\tikzstyle{every state}[align=center,fill=none,draw=black,thick,text=black,minimum size=0pt]
\tikzstyle{every node}=[font=\small]
%,minimum size=0.02cm]

\node[state](A0){};
\draw (A0) node[left,xshift = -0.5cm] {$\cA$} ;
\node[state](A1)[right of=A0]{};
\node[state](A2)[right of=A1]{};
\node[state,fill=gray](A3)[right of=A2]{};
\node[state](A4)[right of=A3]{};
\node[state](A5)[right of=A4]{};
\node[state](A6)[right of=A4]{};
%\node[state,fill=black](A7)[right of=A6,yshift=-0.25cm]{};

\node[state](B0)[below of=A0,yshift=0.25cm]{};
\draw (B0) node[left,xshift = -0.5cm] {$\cB$} ;
\node[state](B1)[below of=A1,yshift=0.25cm]{};
\node[state](B2)[below of=A2,yshift=0.25cm]{};
\node[state,fill=gray](B3)[below of=A3,yshift=0.25cm]{};
\node[state](B4)[below of=A4,yshift=0.25cm]{};
\node[state](B5)[below of=A5,yshift=0.25cm]{};
\node[state](B6)[below of=A6,yshift=0.25cm]{};
%\node[state](B7)[below of=A7]{};

\node[state](C0)[below of=B0,yshift=0.25cm]{};
\draw (C0) node[left,xshift = -0.5cm] {$\cC$} ;
\node[state](C1)[below of=B1,yshift=0.25cm]{};
\node[state](C2)[below of=B2,yshift=0.25cm]{};
\node[state,fill=gray](C3)[below of=B3,yshift=0.25cm]{};
\node[state](C4)[below of=B4,yshift=0.25cm]{};
\node[state](C5)[below of=B5,yshift=0.25cm]{};
\node[state](C6)[below of=B6,yshift=0.25cm]{};
%\node[state](C7)[below of=B7]{};

\node[state](D0)[below of=C0,yshift=0.25cm]{};
\draw (D0) node[left,xshift = -0.5cm] {$\cD$} ;
\node[state](D1)[below of=C1,yshift=0.25cm]{};
\node[state](D2)[below of=C2,yshift=0.25cm]{};
\node[state,fill=gray](D3)[below of=C3,yshift=0.25cm]{};
\node[state](D4)[below of=C4,yshift=0.25cm]{};
\node[state](D5)[below of=C5,yshift=0.25cm]{};
\node[state](D6)[below of=C6,yshift=0.25cm]{};
%\node[state](D7)[below of=C7]{};

\node[state](X0)[below of=D0,yshift=0.25cm]{};
\draw (X0) node[left,xshift = -0.5cm] {$\cX^{(1)}$} ;
\node[state](X1)[below of=D1,yshift=0.25cm]{};
\node[state](X2)[below of=D2,yshift=0.25cm]{};
\node[state,fill=gray](X3)[below of=D3,yshift=0.25cm]{};
\node[state](X4)[below of=D4,yshift=0.25cm]{};
\node[state](X5)[below of=D5,yshift=0.25cm]{};
\node[state](X6)[below of=D6,yshift=0.25cm]{};
%\node[state](X7)[below of=D7]{};

\node[state](Y0)[below of=X0,yshift=0.25cm]{};
\draw (Y0) node[left,xshift = -0.5cm] {$\cX^{(2)}$} ;
\node[state](Y1)[below of=X1,yshift=0.25cm]{};
\node[state](Y2)[below of=X2,yshift=0.25cm]{};
\node[state,fill=gray](Y3)[below of=X3,yshift=0.25cm]{};
\node[state](Y4)[below of=X4,yshift=0.25cm]{};
\node[state](Y5)[below of=X5,yshift=0.25cm]{};
\node[state](Y6)[below of=X6,yshift=0.25cm]{};
%\node[state](Y7)[below of=X7]{};

\node[state,fill=black](P0)[below of=Y0,yshift=0.4cm]{};
\node[state,fill=black](P1)[below of=Y1,yshift=0.4cm]{};
\node[state,fill=black](P2)[below of=Y2,yshift=0.4cm]{};
\node[state,fill=black](P3)[below of=Y3,yshift=0.4cm]{};
\node[state,fill=black](P4)[below of=Y4,yshift=0.4cm]{};
\node[state,fill=black](P5)[below of=Y5,yshift=0.4cm]{};
\node[state,fill=black](P6)[below of=Y6,yshift=0.4cm]{};
%\node[state](Y7)[below of=X7]{};

\path(A3)edge [bend left=30]  node {}(B4);
\path(A3)edge [bend right=30] node {}(X3);

\path(B3)edge [bend left] node {}(B4);
\path(B3)edge [bend left=45] node {}(X3);

\path(C3)edge [bend left=0] node {}(A4);
\path(C3)edge [bend right=30] node {}(X3);

\path(D3)edge [bend left=0] node {}(A4);
\path(D3)edge [bend left=30] node {}(X3);

\path(X3)edge [bend left] node {}(Y3);
\path(X3)edge [bend left=15] node {}(C2);

\path(Y3)edge [loop below,min distance=0.6cm] node {}(Y3);
\path(Y3)edge [bend left=15] node {}(D2);

\draw[dashed,-] (-0.35,-5.85) -- (7,-5.85);
\draw[dashed,-] (-1.35,-5.5) -- (-0.35,-5.5);
\draw[dashed,-] (-0.35,0.2) -- (-0.35,-6.35);
\coordinate [label=left:$T_{n-1}$] (Tn) at (-0.35,-6);
\coordinate [label=left:\small{Clock positions}] (ckp) at (4.5,-6.4);

\end{tikzpicture}

\caption[Markov model for two bit ISI case]
{First order Markov chain model (with extended state space) for two 
bit ISI case. Indicated clock positions and transitions correspond to 
the sub-window $\cW_{\cB-\cC}$. Transitions are indicated for only one 
clock position for brevity. $T_{n-1}$ indicates location of immediate 
previous transition. $\cX^{(i)}$, $i=1,2$, indicates no data transition 
in the previous cycle and in the previous two cycles respectively. 
All transitions out of a state have equal probability.}
\label{fig:ch6:markov-m2}
\end{figure}
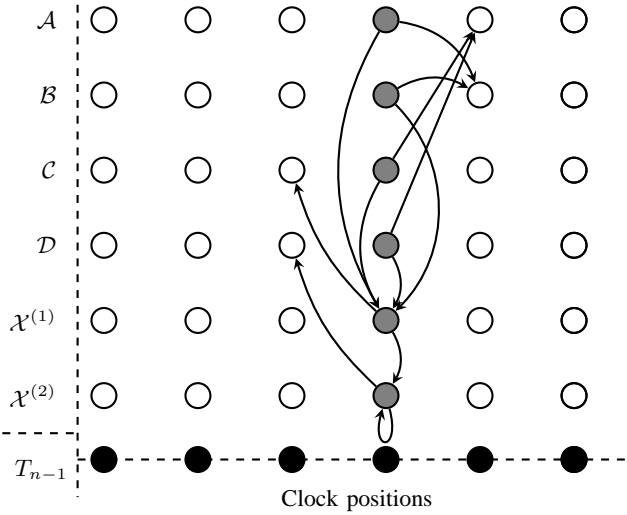

The second order Markov chain model for this system is shown in 
Fig.~\ref{fig:ch6:markov-m2}. To capture the memory of the system, six 
states are used for each clock position. Four of these states 
correspond to the previous data transition ($T_{n-1}$) occurring at 
$\cA$, $\cB$, $\cC$ and $\cD$. The remaining two states are 
i) $\cX^{(1)}$ which indicates no data transition in the previous 
cycle, and ii) $\cX^{(2)}$ which indicates no data transition in the 
previous two cycles. 

For example, when $T_{n-1}=\cA$, the source can be in either state 
$S_2$ or $S_5$ (refer Fig.~\ref{fig:ch6:twobit_fsm}) and the next 
cycle will either not have a data transition or have a data transition 
at position $\cB$. When the initial clock is in the sub-window 
$\cW_{\cB-\cC}$, a data transition at $\cB$ results in a clock shift to 
the right by 1$\delta$. Thus, the transitions from state at 
$T_{n-1}=\cA$ are either to state $\cX^{(1)}$ at the same clock 
position or to state $T_{n-1}=\cB$ of the next clock position to the 
right. Extending the analysis to the rest of the FSM in 
Fig.~\ref{fig:ch6:twobit_fsm}, the rest of the state transitions can 
be determined. In Fig.~\ref{fig:ch6:markov-m2}, the transitions are 
shown for only one clock position for brevity, in particular, for the 
clock positions in the sub-window $\cW_{\cB-\cC}$. Note that all the 
state transitions have a probability of 0.5 when the data source 
outputs 0 and 1 with equal probability.

\begin{figure}[h!]
\centering
\psfrag{Behavioural}{\scriptsize{Behavioural}}
\psfrag{simulation}{\scriptsize{Simulation}}
\psfrag{Model}{\scriptsize{Model}}
\psfrag{A}{\small{$\cA$}}
\psfrag{B}{\small{$\cB$}}
\psfrag{C}{\small{$\cC$}}
\psfrag{D}{\small{$\cD$}}
\psfrag{prediction}{\scriptsize{prediction}}
\psfrag{Initial clock position}{\small{Initial clock position}}
\psfrag{Absorbtion Time (#cycles)}{\small{Mean absorption time (cycles)}}
\psfrag{Tw}{\small{$T_{\cW}2$}}
\psfrag{-Tw}{\small{-$T_{\cW}/2$}}
\includegraphics[width=0.45\columnwidth]{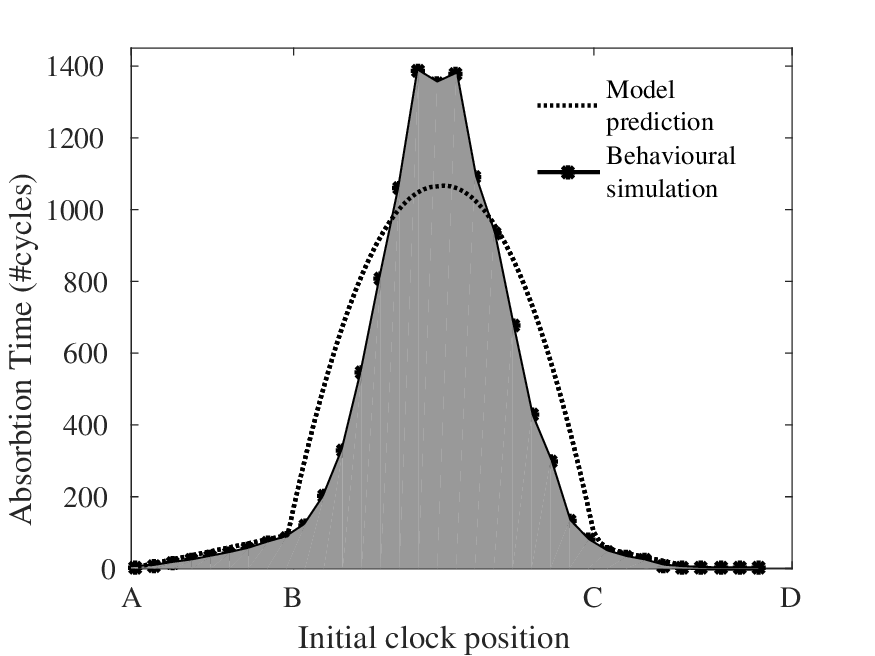}
\caption{Mean absorption time as predicted from the Markov chain model and
from behavioural
simulations. Each data point in the behavioural simulation is the average of
350 runs.}
\label{fig:ch6:model-mean-abs-2b}
\end{figure}

Similar to the 1 bit ISI case, the mean time to absorption for the 2 bit 
ISI case was determined using the Markov chain model. Also, behavioural 
simulations were performed. For the eye diagram shown in 
Fig.~\ref{fig:ch6:high_isi}, the window size is about 30\% of the clock 
period. Here $\cW_{\cA-\cB}\sim 8\%$, $\cW_{\cB-\cC}\sim 15 \%$, and
$\cW_{\cC-\cD}\sim 7 \%$. 
%Accordingly, the Markov chain was 
%constructed and analyzed using a linear equation solver.
The mean time to absorption for each clock position was taken as the 
average of the mean times to absorption from each of the six states 
in the Markov chain that correspond to that position. For the 
behavioural simulations, a 20 section RC interconnect was used with 
a VerilogA behavioural description of the clock retiming circuit. 
In order to reduce the simulation time and the required sample size, 
a step size of 3.5$\tau$ was used for this analysis. The combined plots 
of the mean time to absorption for 2 bit ISI case obtained from 
behavioural simulations and Markov chain model predictions are shown in 
Fig.~\ref{fig:ch6:model-mean-abs-2b}. 
%
\begin{comment}[h!]
\centering
\psfrag{Behavioural}{\scriptsize{Behavioural}}
\psfrag{simulation}{\scriptsize{Simulation}}
\psfrag{Model}{\scriptsize{Model}}
\psfrag{A}{\small{$\cA$}}
\psfrag{B}{\small{$\cB$}}
\psfrag{C}{\small{$\cC$}}
\psfrag{D}{\small{$\cD$}}
\psfrag{prediction}{\scriptsize{prediction}}
\psfrag{Initial clock position}{\small{Initial clock position}}
\psfrag{Absorbtion Time (#cycles)}{\small{Mean absorption time (cycles)}}
\psfrag{Tw}{\small{$T_{\cW}2$}}
\psfrag{-Tw}{\small{-$T_{\cW}/2$}}
\includegraphics[width=0.45\columnwidth]{./figs/two_bit_isi_data_std.eps}
\caption{Standard deviations of absorption time as predicted from 
the Markov chain model and from behavioural simulations. Each data 
point in the behavioural simulation is the average of 350 runs.}
\label{fig:ch6:model-std-abs-2b}
\end{comment}
%
The behavioural simulations used an RC circuit for the interconnect, 
whereas the Markov chain model truncated the ISI to 2 bits.
Note that the peak of the mean time to absorption, as predicted by 
the model, is about 1100 cycles when the step size is 3.5$\tau$. 
%The fit of the model for two bit ISI is som
This can be as high as 12000 cycles when the step size is 1$\tau$.

\textcolor{blue}{
Thus, we can conclude that the predictions of the settling time 
of the clock retiming circuits  by the Markov chain model have 
good accuracy. 
%Thus, we conclude that the predictions by the Markov 
%chain model of the retiming circuit represent the settling time of 
%the clock retiming circuit with good accuracy. 
The Markov chain model predictions are presented for an equiprobable 
bit sequence as the data. % for the computations. 
As seen in Fig.~\ref{fig:ch6:model-mean-abs-2b}, these 
predictions differ slightly from those 
given  by the behavioural simulations, where the latter are 
performed for a bit sequence generated by the random number 
generator in VerilogA. The deviation is owing to the 
following: i) the random bit sequence generated is not exactly 
equiprobable, ii) a finite number of simulations are used.}

%
%The behavioural simulations are performed with a random number 
%generator in VerilogA. The small deviations of the results of the 
%behavioural simulations from the model predictions could be attributed 
%to i) The finite number of simulations used, and ii) the fact that 
%the computer generated random bit streams are never exactly 
%equiprobable.}

\subsection{Markov chain model for Gaussian distributed 
random jitter}
This section discusses the increase in the settling time due to random 
jitter in the clock. There are two sources of clock jitter in digital 
systems: inherent jitter in the clock generating oscillator and jitter 
introduced by the clock distribution network. The former can be modelled 
as a Gaussian distribution~\cite{generated_jitter}. The latter, 
however, depends on several factors and can be random with arbitrary 
distribution~\cite{distribution_jitter}. The analysis in this section 
assumes that the jitter has a Gaussian distribution.

For ease of analysis, it is assumed that the receiver clock is noise 
free and all the jitter is in the transmitter clock. This assumption 
is valid as long as the channel response is constant over the spectrum 
of the noisy clock. The jitter histogram is assumed to be a Gaussian 
distribution with a standard deviation of $\sigma_{ck}$. The sampling 
clock position ($t_{ck}$) with respect to the mean data transition 
position can be represented as $m\tau$, where $m$ is an integer taking 
values from $-\infty$ to $\infty$. Hence, with respect to the sampling 
clock position, the mean of the data transition can be written as 
$-m\tau$. Unlike the analysis for ISI induced jitter where the 
window of susceptibility \gls{cW} was bounded, random jitter may be 
unbounded, and hence, the size of the window \gls{cW} may be arbitrarily 
large. For our analysis, a window size of $\pm 3\sigma$, i.e., 
$|m\tau|<3\sigma_{ck}$, is assumed. This covers $>99.999\%$ of the 
possible data transition positions.

\begin{figure}[h!]
\centering
\psfrag{ndel}{\small{$t_{ck}$}}
\psfrag{0}{\small{\hspace{-3ex}$-m\tau$}}
\psfrag{PDF}{\small{PDF}}
\psfrag{A}{\small{A}}
\includegraphics[width=8cm]{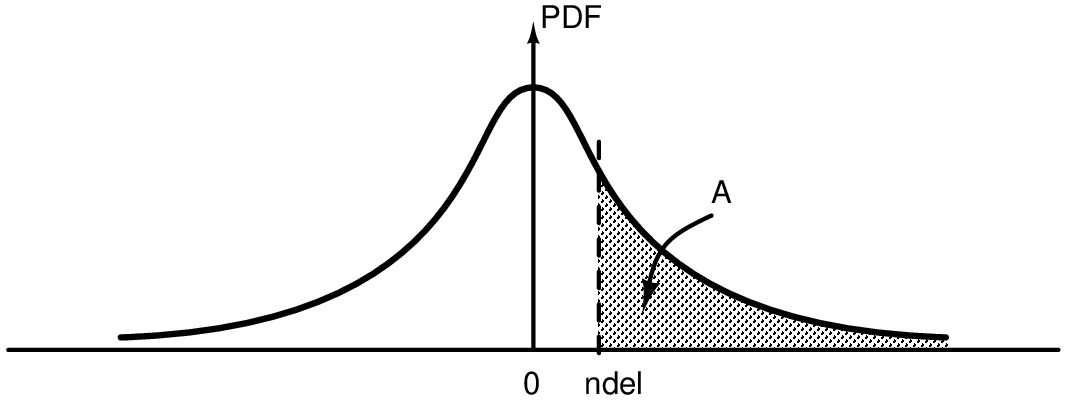}
\caption{Probability distribution of clock transition time and the position 
of data transition for Gaussian random jitter.}
\label{fig:ch6:rand_jitter_pdf}
\end{figure}

\begin{figure}[h!]
\centering
\psfrag{Absorbtion Time (#cycles)}{\small{
\hspace{-5ex}Mean absorption Time (cycles)}}
\psfrag{Standard deviation (#cycles)}{\small{Standard deviation (cycles)}}
\psfrag{s=20d}{\small{$\sigma_{ck}=20\tau $}}
\psfrag{s}{\small{$\sigma_{ck}$}}
\psfrag{2s}{\small{$2\sigma_{ck}$}}
\psfrag{3s}{\small{$3\sigma_{ck} $}}
\psfrag{0}{\small{$0$}}
\psfrag{-s}{\small{$-\sigma_{ck}$}}
\psfrag{-2s}{\small{$-2\sigma_{ck}$}}
\psfrag{-3s}{\small{$-3\sigma_{ck}$}}
\psfrag{Initial clock position}{\small{Initial data position}}
\includegraphics[width=0.475\columnwidth]{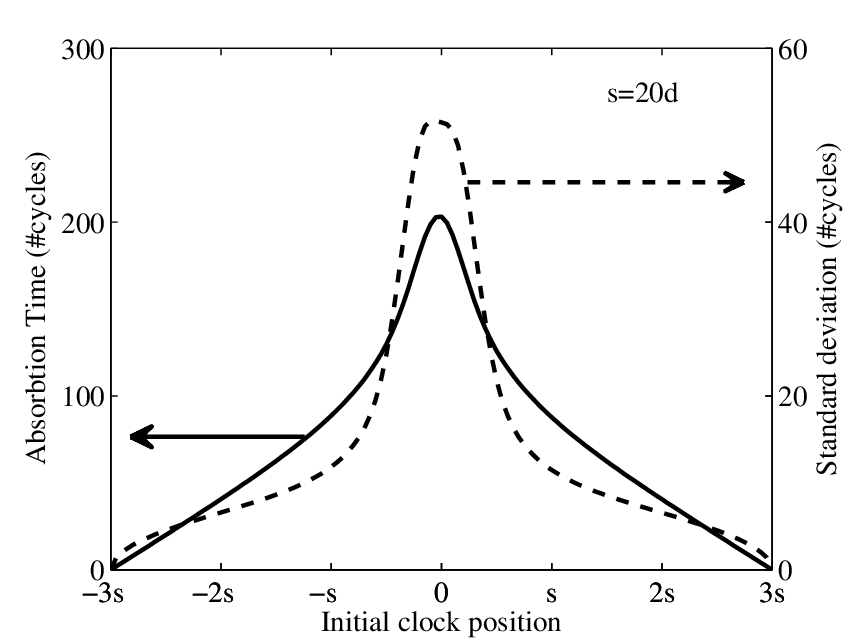}
\caption{Mean absorption time and standard deviation of the 
absorption time with Gaussian distributed random jitter. The standard 
deviation of the clock jitter is assumed to be $20\tau$ (i.e., 0.02UI).}
\label{fig:ch6:model-random-jitter}
\end{figure}

When the clock transition is to the right of the mean data transition 
position ($m>0$), the circuit shifts the clock to the right towards final 
lock ($m$ is incremented). 
However, even when $m>0$, the random jitter in the data can cause the 
instantaneous data transition to occur to the right of the sampling clock  
position. This results in a clock shift in the wrong direction.
Fig.~\ref{fig:ch6:rand_jitter_pdf} shows the probability 
distribution of the clock transition time and the data transition position. 
The shaded area represents the probability that the data transition 
occurs to the right of the sampling clock position, when the mean of 
the data transition is to the left of the sampling clock transition 
($m>0$). This is the probability of a phase update in the wrong 
direction. This probability can be calculated as follows.
\begin{align*}
\prob(\text{update is wrong}) &= \int_{t_{ck}}^{\infty}\frac{e^\frac{-(x+m\tau)^2}{2\sigma_{ck}^2}}{\sqrt{2\pi}\sigma_{ck}}dx  \\
%&= 1-\int_{-\infty}^{m\tau}\frac{e^\frac{-(x+m\tau)^2}{2\sigma_{ck}^2}}{\sqrt{2\pi}\sigma_{ck}}dx \\
%&= 1-\frac{1}{2}\left[1+erf\left(\frac{2m\tau}{\sqrt{2}\sigma_{ck}}\right)\right]  \\
%&= \frac{1}{2} - \frac{1}{2}erf\left(\frac{2m\tau}{\sqrt{2}\sigma_{ck}}\right),
&= \frac{1}{2}\mbox{\it erfc}\left(\frac{2m\tau}{\sqrt{2}\sigma_{ck}}\right),
\end{align*}
where $\mbox{\it erfc}(x)$ is the standard complementary  error 
function~\cite{markov}. The probability of an update in the correct 
direction can then be calculated as 
\begin{align*}
\prob(\text{update is right})=1-\prob(\text{update is wrong}). 
\end{align*}
Note that these probability values will be scaled by a factor which 
equals the probability of transitions since the transitions occur 
independent of the clock jitter.
%Note that these probability values will be scaled by the probability of 
%transitions. 
For a source which outputs 1 and 0 with equal probability, 
the probability of transition is also $1/2$. 
%Since the data transitions 
%are independent of the clock jitter, the probabilities can be 
%multiplied. 
This equation is then used to construct the probability 
transition matrix of the Markov chain to predict the mean time to 
absorption and the standard deviation of the time to absorption. As 
noted earlier, absorption in this case is defined as the phase 
difference between the mean of data position and sampling clock 
position being $>3\sigma _{ck}$. Fig.~\ref{fig:ch6:model-random-jitter} 
shows the plots of the mean time to absorption and the standard 
deviation of the time to absorption with the initial clock position 
as estimated by the Markov chain model, when 
%the standard deviation of the clock jitter is 
$\sigma _{ck}=20\tau$, i.e., standard deviation of the clock jitter is 
2\% of the clock period.

\subsection{Modeling the effect of ISI and random jitter}
The Markov chain model of the system can be easily extended to analyze 
the combined effect of data dependent and random jitter. An example 
case of jitter induced by 1 bit \gls{isi} and Gaussian distributed 
random jitter is shown in this section. When the jitter is solely due 
to 1 bit \gls{isi}, the data follows one of two distinct traces as 
discussed in Section~\ref{subsec:ch6:markovmodel-1bit}. When random 
jitter is also present, the zero crossing of the data spreads around 
these two zero crossing positions. Since the random jitter is 
independent of the data, the probability distribution of the data zero 
crossing positions is equal to the product of the probability 
distribution of the zero crossing positions due to random jitter and 
due to 1 bit ISI.
%contributions of each of the sources of jitter. 
For the purpose of analysis it assumed that the 
receiver clock is jitter free and the transmitter clock has all the 
random jitter and the channel introduces data dependent jitter.

\begin{figure}[h!]
\centering
\psfrag{tck}{\small{$t_{ck}$}}
\psfrag{0}{\small{\hspace{-3ex}$-m\tau$}}
\psfrag{B}{\small{$\cB$}}
\psfrag{A}{\small{$\cA$}}
\psfrag{+3s}{\small{$3\sigma_{ck}$}}
\psfrag{-3s}{\small{$-3\sigma_{ck}$}}
\includegraphics[width=0.5\columnwidth]{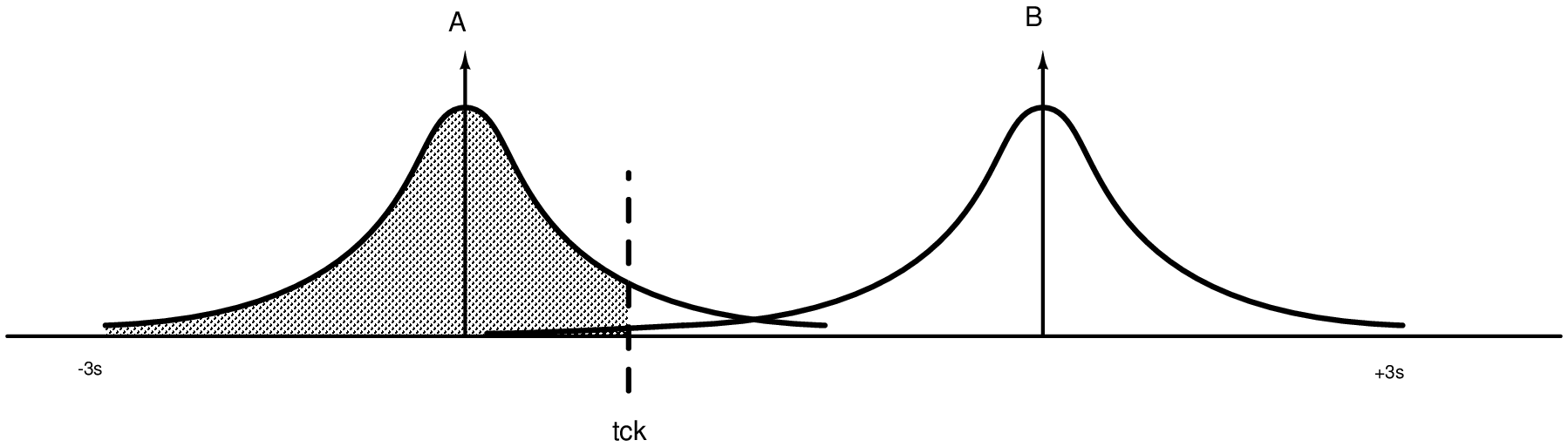}
\caption{Probability distribution of clock transition time and the 
position of data transition for Gaussian random jitter along with 1 bit ISI.}
\label{fig:ch6:combined_jitter}
\end{figure}

Fig.~\ref{fig:ch6:combined_jitter} shows the distribution of the 
data zero crossing positions and the sampling clock position when the data 
signal has both random jitter and 1 bit \gls{isi} induced jitter. The 
window of susceptibility is now widened to 
$3\sigma _{ck} + \cW_{\cA-\cB} + 3\sigma _{ck}$.

The effect of the data jitter can be modelled as a shift in the mean of the Gaussian random jitter. Hence, for 
random jitter along with jitter due to 1 bit \gls{isi}, the jitter at any time can be considered to be Gaussian 
distributed, with a standard deviation of $\sigma_{ck}$ and mean at either $\cA$ or $\cB$.
Consider an example clock position ($t_{ck}$) as shown in 
Fig.~\ref{fig:ch6:combined_jitter}. Choosing $\cA$ as the 
reference position for calculations, the probability of a data 
transition occurring to the left of the clock is given by
\begin{align*}
\prob(T_L) = \begin{aligned}[t]
& \prob(\cA)\cdot\Phi\left(\frac{t_{ck} - 0}{\sigma_{ck}}\right) 
 + \prob(\cB)\cdot\Phi \left(\frac{t_{ck} - \cW_{\cA - \cB}}{\sigma_{ck}}\right),
\end{aligned} 
\end{align*}
where $\Phi\left(\frac{x-\mu}{\sigma}\right)$ is the \gls{cdf} of a general Gaussian distribution with 
mean $\mu$ and standard deviation $\sigma$.
The probability of a data transition to the right of the clock can then be calculated as 
\begin{align*}
\prob(T_R) = 1- \prob(T_N) - \prob(T_L),
\end{align*}
where $\prob(T_N)$ is the probability of having no data transition 
in that clock cycle.

\begin{figure}[h!]
\centering
\psfrag{Absorbtion Time (#cycles)}{\small{
\hspace{-5ex}Mean absorption Time (cycles)}}
\psfrag{Standard deviation (#cycles)}{\small{Standard deviation (cycles)}}
\psfrag{s=20d}{\small{$\sigma_{ck}=20\tau $}}
\psfrag{s}{\small{$\sigma_{ck}$}}
\psfrag{2s}{\small{$2\sigma_{ck}$}}
\psfrag{3s}{\small{$3\sigma_{ck} $}}
\psfrag{0}{\small{$0$}}
\psfrag{A}{\scriptsize{$\cA$}}
\psfrag{B}{\scriptsize{$\cB$}}

\psfrag{-s}{\small{$-\sigma_{ck}$}}
\psfrag{-2s}{\small{$-2\sigma_{ck}$}}
\psfrag{-3s}{\small{$-3\sigma_{ck}$}}
\psfrag{Initial clock position}{\small{Initial data position}}
\includegraphics[width=0.45\columnwidth]{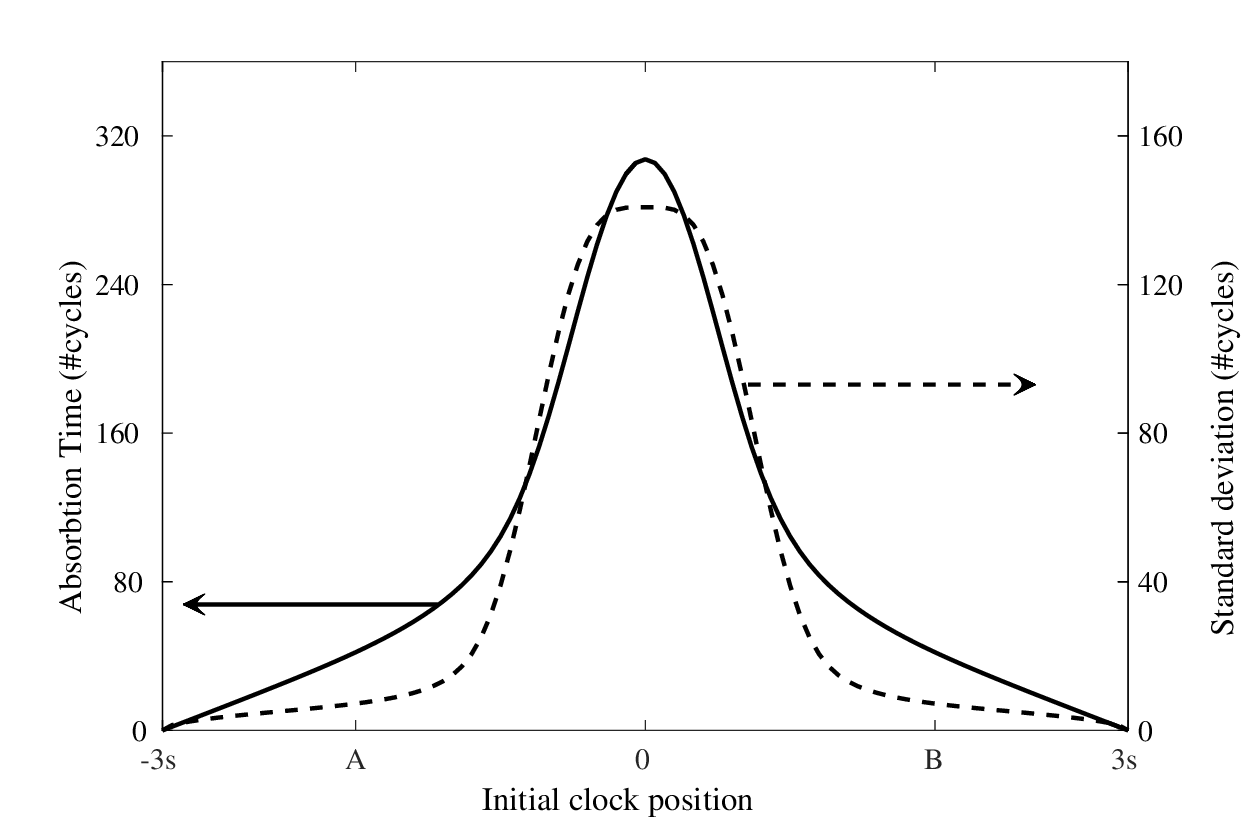}
\caption{Mean absorption time and standard deviation of the
absorption time when data has 1 bit data dependent jitter and Gaussian 
distributed random jitter. The standard
deviation of the random clock jitter is assumed to be $10\tau$ (i.e., 0.01UI).}
\label{fig:ch6:model-combined-jitter}
\end{figure}

The Markov chain along with its transition probabilities 
was determined in this manner.
Fig.~\ref{fig:ch6:model-combined-jitter} shows 
the mean absorption time and the standard deviation of the absorption 
time obtained for data with timing jitter induced by 1 bit \gls{isi} in addition 
to Gaussian distributed random jitter with standard deviation of 1\% of the 
clock period.

%% file: quantification.tex
\section{Bounds on the settling time}
\label{sec:quant}
As the settling time is a random quantity, it is not possible to 
determine 
it exactly. However, in this section, we determine bounds on the 
settling time for a given value of probability of absorption, i.e.,
confidence level. Toward this we leverage the fact that the settling 
time of the circuit is maximum when the initial clock is
at the center of the window of susceptibility (for a system without 
any programmed bias). In particular, the bound on the mean settling 
time will equal this maximum value of settling time. 

Let $\mu_{(0)}$ be the row vector corresponding to the initial 
probability mass function on the states of the Markov chain. 
Here $\mu_{(0)}$ contains exactly one entry with value 1 which 
corresponds to the initial clock position (at the center of the 
window~$\cW$), while the entries 
corresponding to the rest of positions are zeroes. Let the 
probability transition matrix of this Markov chain be 
denoted by $\bar{P}$. Using standard results for Markov 
chains~\cite{markov}, we know that given $\mu_{(0)}$, the probability 
mass function on the states of the Markov chain after $n$ transitions 
is given by 

\begin{align*}
\mu_{(n)} = \mu_{(0)} \cdot \bar{P}^n .
\end{align*}
For a Markov chain with absorbing states, the probability of absorption 
after $n$ transitions equals the sum of those entries of $\mu_{(n)}$ that 
correspond 
to the absorbing states. Hence, one can iteratively find the number of 
transitions required for absorption to occur for a given absorption 
confidence level (i.e., probability of absorption).

\begin{figure}[h!]
\centering
\psfrag{n = Number of transitions}{\small{Number of transitions ($n$).}}
\psfrag{P absorption after n cycles}{\small{\hspace{-3ex}Prob. of 
absorption after $n$ cycles}}
\includegraphics[width=0.45\columnwidth]{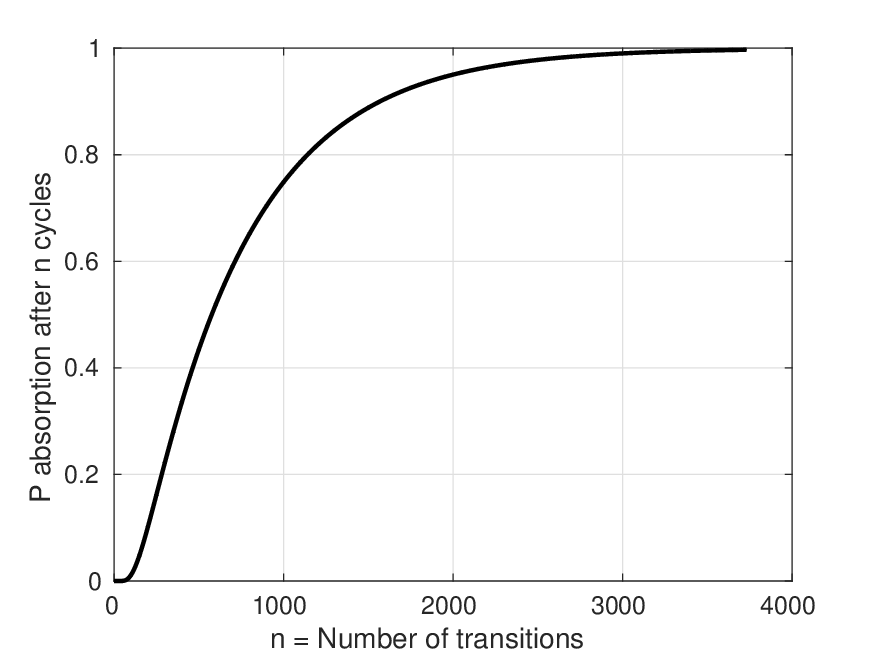}
\caption{Probability of absorption as a function of the number of 
transitions.}
\label{fig:ch6:cabs}
\end{figure}

\begin{figure}[h!]
\centering
\psfrag{n = Number of transitions}{\small{Number of transitions ($n$).}}
\psfrag{P absorption in exactly n cycles}{\small{\hspace{-3ex}Prob. of absorption
in exactly $n$ cycles}}
\includegraphics[width=0.45\columnwidth]{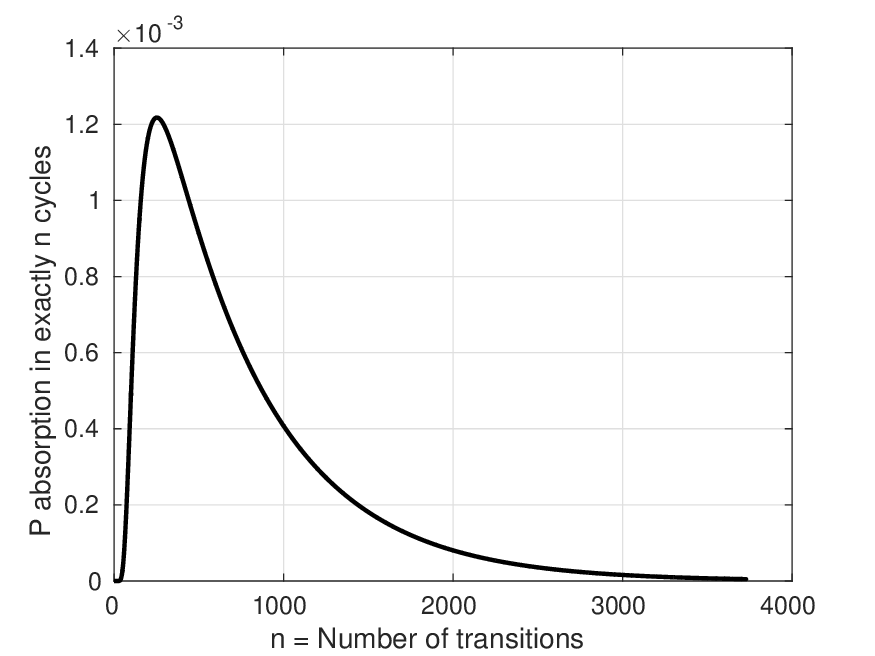}
\caption{Probability of absorption in exactly $n$ transitions as 
a function of $n$. Here the size of the window $\cW$ is 40 steps.}
\label{fig:ch6:pabs}
\end{figure}

For illustrative purposes, we present results for a 1 bit ISI case 
with a window of susceptibility of size 40 steps. From the procedure 
described earlier to obtain a bound on the settling time, we now
calculate the probability of absorption of the Markov chain as a 
function of the number of transitions. The results obtained are 
shown in Fig.~\ref{fig:ch6:cabs}. 
In fact, this plot corresponds to the cumulative distribution 
function of the absorption time. Hence, it directly follows that the 
probability of absorption in exactly $n$ cycles corresponds to the 
slope of the above graph. This is shown in 
Fig.~\ref{fig:ch6:pabs}. 
Observe that the distribution of probability of absorption is skewed 
as one would expect. The probability of absorption is zero for 
$n$ less than the distance to the nearest absorbing state. 
Also, for very large values of $n$ this probability approaches zero.
%i.e., the probability of absorption is asymptotically.
%The graph then approaches zero asymptot
% For number of transitions less than the 
%distance to the absorbing state is always zero.

\begin{figure}[h!]
\centering
\psfrag{Window size in number of steps}{\small{Window size in
number of steps}}
\psfrag{n cycles for P abs = 0.99}{\small{\hspace{-6ex}$n$ cycles 
for abs. with $>$99\% confidence}}
%Prob. of
%abs. = 0.99}}
\includegraphics[width=0.45\columnwidth]{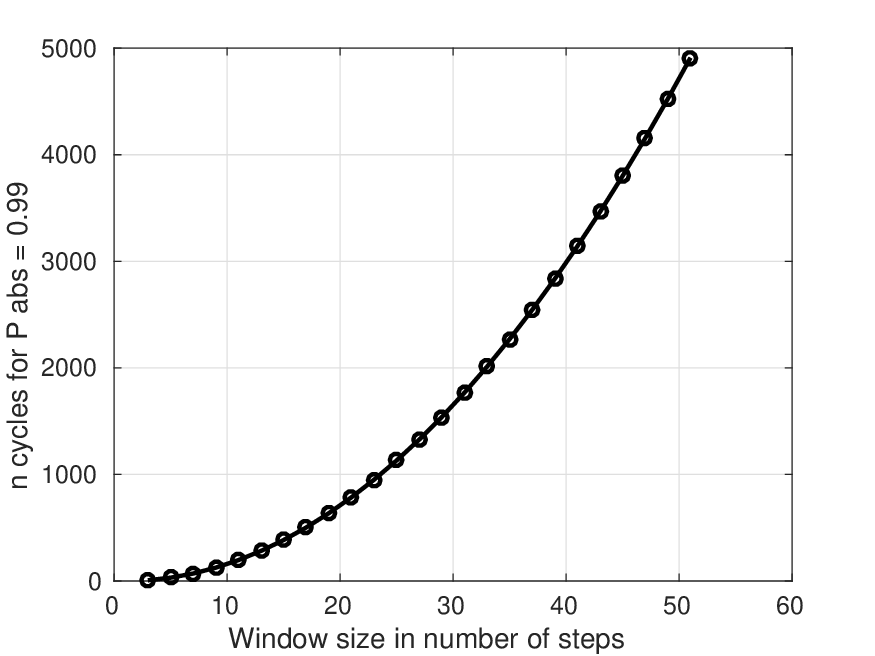}
\caption{Number of cycles $n$ for 
%Probability of 
absorption in $\le n$ transitions, with $>$99\% 
confidence, as a function of window size for 1 bit ISI example.}
\label{fig:ch6:tabs}
\end{figure}

As seen from Fig.~\ref{fig:ch6:cabs}, for absorption with $>$99\% 
surety, 
the Markov chain needs about 3000 state transitions. This number 
depends on the gain of the system and increasing the gain reduces the 
window size (in terms of number of phase steps), and hence, the settling 
time. An analysis of the number of 
transitions needed for absorption with $>$99\% confidence with the size of 
the window, in terms of number of steps, was done. 
Fig.~\ref{fig:ch6:tabs} shows the plot obtained from 
this analysis. 
%This 
%As discussed earlier, the window size can be controlled by changing the gain of the 
%loop. 

%% file: reduction.tex
\section{Techniques for reducing the settling time}
\label{sec:reduce}
\subsection{The coarse first synchronizer}
%
\begin{comment}
\begin{figure*}[h!]
\centering
\psfrag{CkRx}{\small{$\phi_{Rx}$}}
\psfrag{Ckd}{\small{$\phi_d$}}
\psfrag{Data}{\small{Data}}
\psfrag{Input}{\small{Input}}
\psfrag{Phase}{\small{Phase}}
\psfrag{Detector}{\small{Detector}}
\psfrag{Charge}{\small{Charge}}
\psfrag{Pump}{\small{Pump}}
\psfrag{(weak)}{\small{(weak)}}
\psfrag{(strong)}{\small{(strong)}}
\psfrag{Logic}{\small{Logic}}
\psfrag{UPDN}{\footnotesize{UP DN}}
\psfrag{UP}{\footnotesize{$UP$}}
\psfrag{UPst}{\footnotesize{$UP_{strong}$}}
\psfrag{DNst}{\footnotesize{$DN_{strong}$}}
\psfrag{DN}{\footnotesize{$DN$}}
\psfrag{Counter}{\footnotesize{Counter}}
\psfrag{Divider}{\small{Divider}}
\psfrag{Switch Matrix}{\small{Switch Matrix}}
\psfrag{DLL}{\small{DLL}}
\psfrag{VCDL}{\small{VCDL}}
\psfrag{Window}{\small{Window}}
\psfrag{Comparator}{\small{Comparator}}
\psfrag{Retimed Data}{\small{Retimed Data}}
\psfrag{UD}{\footnotesize{$\overline{UP}/DN$}}
\psfrag{EN}{\footnotesize{$Enable$}}
\psfrag{VC}{\small{$V_{c}$}}
\psfrag{Qs}{\footnotesize{$Q_0-Q_9$}}
\psfrag{Fine}{\small{Fine tuning loop}}
\psfrag{Coarse}{\small{Coarse tuning loop}}
\psfrag{phin}{\small{$ $}}
\includegraphics[width=14cm]{./figs/cpa-coarse_first.eps}
\caption{Block diagram of the clock synchronizer that used coarse and fine 
correction for achieving lock, with coarse tuning preceding the fine 
synchronizer.}
\label{fig:ch6:coarse-first}
\end{figure*}
\end{comment}

The settling time of the circuit depends on the loop gain or the step 
size of the phase correction circuit. A large step size will mean that 
the size of the window of susceptibility (in term of number of steps) 
is small. This results in quick settling, as was discussed in the 
previous section. However, a large step size results in high jitter in 
the clock after lock is achieved. The synchronizer proposed 
in~\cite{naveen_vlsi17} uses a coarse and a fine correction loop for 
accurate synchronization. The circuit designed there first 
tries to achieve lock using the fine tuning loop. If lock cannot be 
achieved in the limited range of the fine tuning loop, a coarse 
correction is initiated. If the fine tuning loop is disabled 
during the initial settling period, it enables us to have a much larger
step size, and hence, a much faster settling. For instance, when a 
10 phase DLL is used, and even if a horizontal eye opening of 50\% is 
available, the size of $\cW$ is only 5 steps.

\begin{figure}[h!]
\centering
\psfrag{Weak charge pump}{\scriptsize{Weak charge pump}}
\psfrag{charge pump}{\scriptsize{charge pump}}
\psfrag{Strong}{\scriptsize{Strong}}
\psfrag{UPS}{\scriptsize{$UP_{strong}$}}
\psfrag{DNS}{\scriptsize{$DN_{strong}$}}
\psfrag{Vbp}{\scriptsize{$V_{bp}$}}
\psfrag{Vbn}{\scriptsize{$V_{bn}$}}
\psfrag{VC}{\scriptsize{$V_c$}}
\psfrag{UP}{\scriptsize{$UP$}}
\psfrag{UPB}{\scriptsize{$\overline{UP}$}}
\psfrag{DN}{\scriptsize{$DN$}}
\psfrag{DNB}{\scriptsize{$\overline{DN}$}}
\includegraphics[width=0.4\columnwidth]{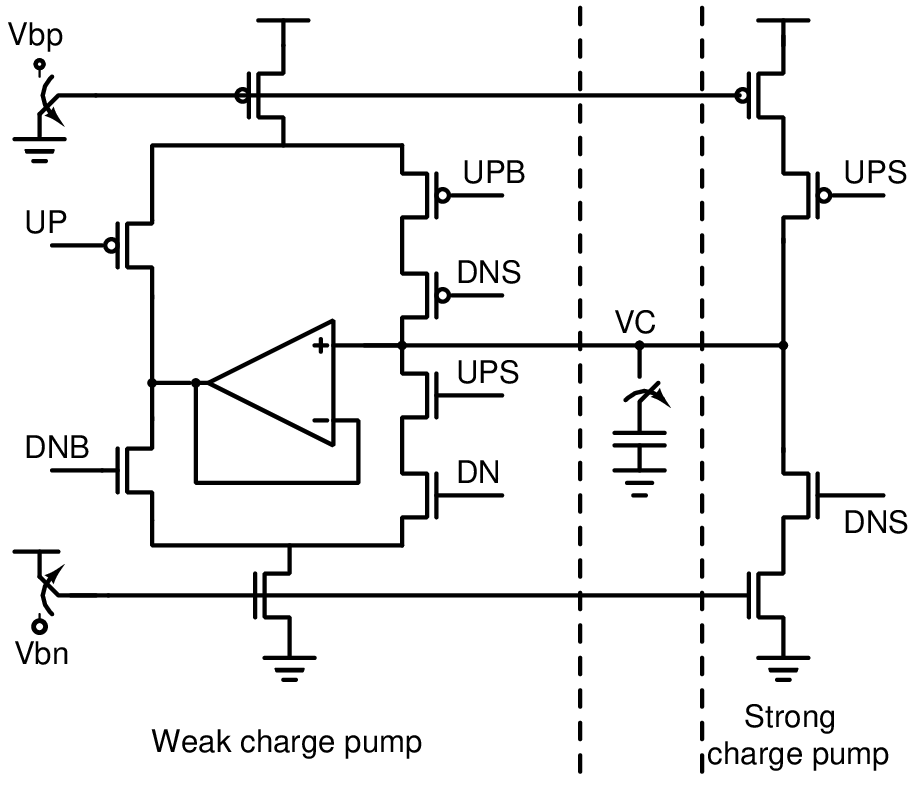}
\caption{Charge pump circuit for enabling coarse first operation of the 
synchronizer. $UP$, $\overline{UP}$, $DN$ and $\overline{DN}$ are driven 
by the fine tuning loop. $UP_{strong}$ and $DN_{strong}$ are 
driven by the coarse tuning loop. $V_{bn}$ and $V_{bp}$ are the bias 
voltages for the NMOS and PMOS current sink and source respectively. The
switches are used to enable the coarse-first mode.}
\label{fig:ch6:coarse-first}
\end{figure}

Fig.~\ref{fig:ch6:coarse-first} shows the charge pump circuit for the 
synchronizer in Fig.~\ref{fig:cdr-coarse-fine}, appropriately modified 
to have a coarse only mode.
%block diagram of the coarse first synchronizer. 
%
Here, during the initial period after startup, a switch 
disconnects the loop filter capacitor and another set of switches 
convert the charge pump to a combinational circuit~\cite{naveen-date16},
i.e., connecting the gate of the PMOS current source to GND and 
gate of the NMOS current sink to VDD.
\textcolor{blue}{This converts the current sources in the fine charge pump to ON switches. This essentially increases the current injected at 
every fine correction step, and the control voltage 
overshoots the bounds set by $V_H$ and $V_L$ in every cycle. 
Hence, the circuit takes a coarse 
correction step at every cycle of the divided clock. This is 
equivalent to 
%This essentially 
disabling the fine tuning loop and ensuring a coarse tuning step 
in every cycle of the 
divided clock that drives the coarse tuning loop. 
After lock is achieved, the loop filter capacitor is connected 
back to the charge pump, and the gates of the PMOS current source and 
NMOS current sink are connected to $V_{bp}$ and $V_{bn}$ respectively, 
to resume normal operation.}
 
The amount of time for which the coarse tuning loop should be run 
before enabling the fine tuning loop depends on the size of $\cW$ and
on the desired confidence level for achieving lock within this period. 
In this example given in Fig.~\ref{fig:cdr-coarse-fine}, the coarse 
synchronizer used a 10 phase (DLL). Assuming an extreme condition 
that the horizontal eye opening is only 50\%, it means that the 
window size is 5 steps. From the analysis presented in the previous 
section, an absorption with $>$99\% confidence level needs 32 cycles 
to be absorbed. This corresponds to 256~ns in absolute time if the 
coarse controller operates on a divide by 16 clock for a 2~GHz 
system clock.
  
\begin{figure}[h!]
\centering
\psfrag{Time}{\small{Two UI's}}
\psfrag{data}{\small{\hspace{-4ex}Receiver i/p (mV)}}
\includegraphics[width=8.5cm]{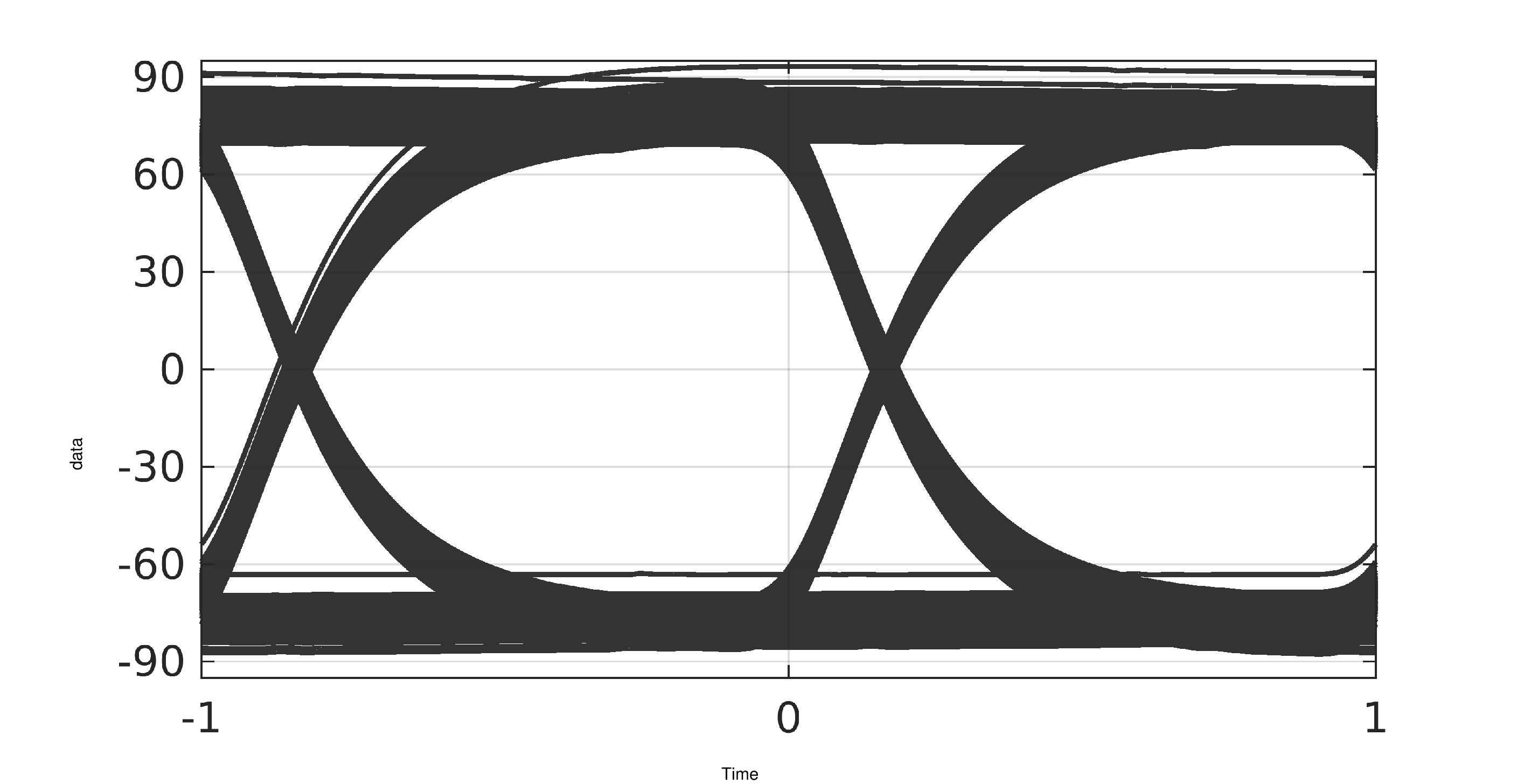}
\caption{Receiver input eye diagram for a benign channel.}
\label{fig:ch6:coarse-eye-low}
\end{figure}

\begin{figure}[h!]
\centering
\psfrag{Time (mus)}{\scriptsize{Time ($\mu$s)}}
\psfrag{Coarse only}{\scriptsize{\hspace{-2ex}Coarse enable}}
\psfrag{Voltage}{\scriptsize{\hspace{-6ex}Control voltage (V)}}
\includegraphics[width=0.45\columnwidth]{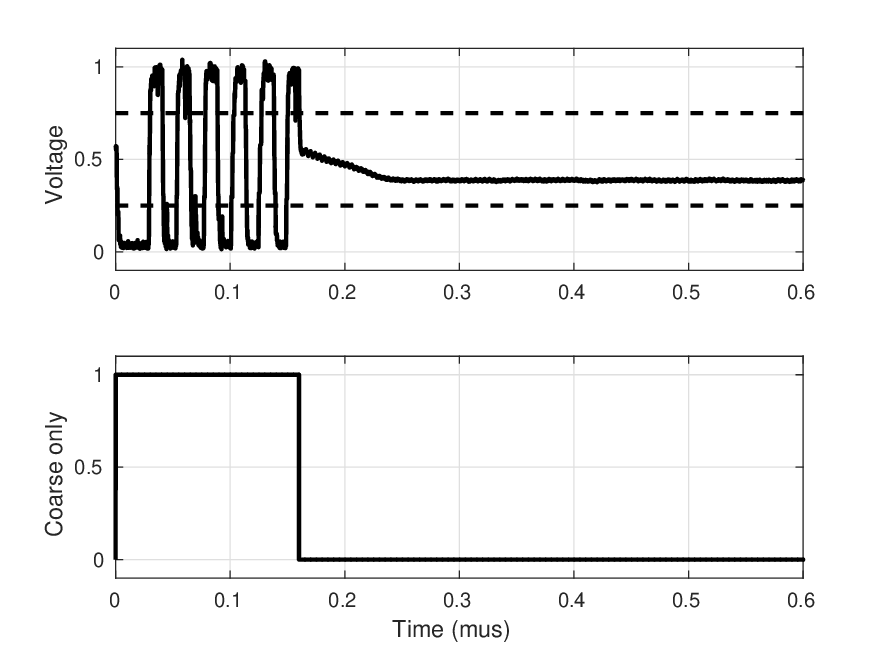}
\caption{Settling of the coarse first synchronizer with input having 
jitter from a benign channel. The lower waveform is the control signal 
for enabling the coarse mode.}
\label{fig:ch6:coarse-first-lowisi}
\end{figure}

The coarse first synchronizer was designed and simulated for a benign 
channel. The receiver input eye diagram for this test is shown in 
Fig.~\ref{fig:ch6:coarse-eye-low}.
For an initial clock phase at the center of $\cW,$ the settling 
behaviour is shown in Fig.~\ref{fig:ch6:coarse-first-lowisi}.
As seen, the circuit escapes the window $\cW$ very fast and settles 
accurately once the fine tuning loop is enabled. The lower waveform is 
the select signal for switching from coarse  mode to normal mode, 
which is run for 160~ns in coarse mode. This signal can be generated 
using a power on reset circuit.

\begin{figure}[h!]
\centering
\psfrag{Time}{\small{Two UI's}}
\psfrag{data}{\small{\hspace{-2ex}Receiver i/p (mV)}}
\includegraphics[width=8.5cm]{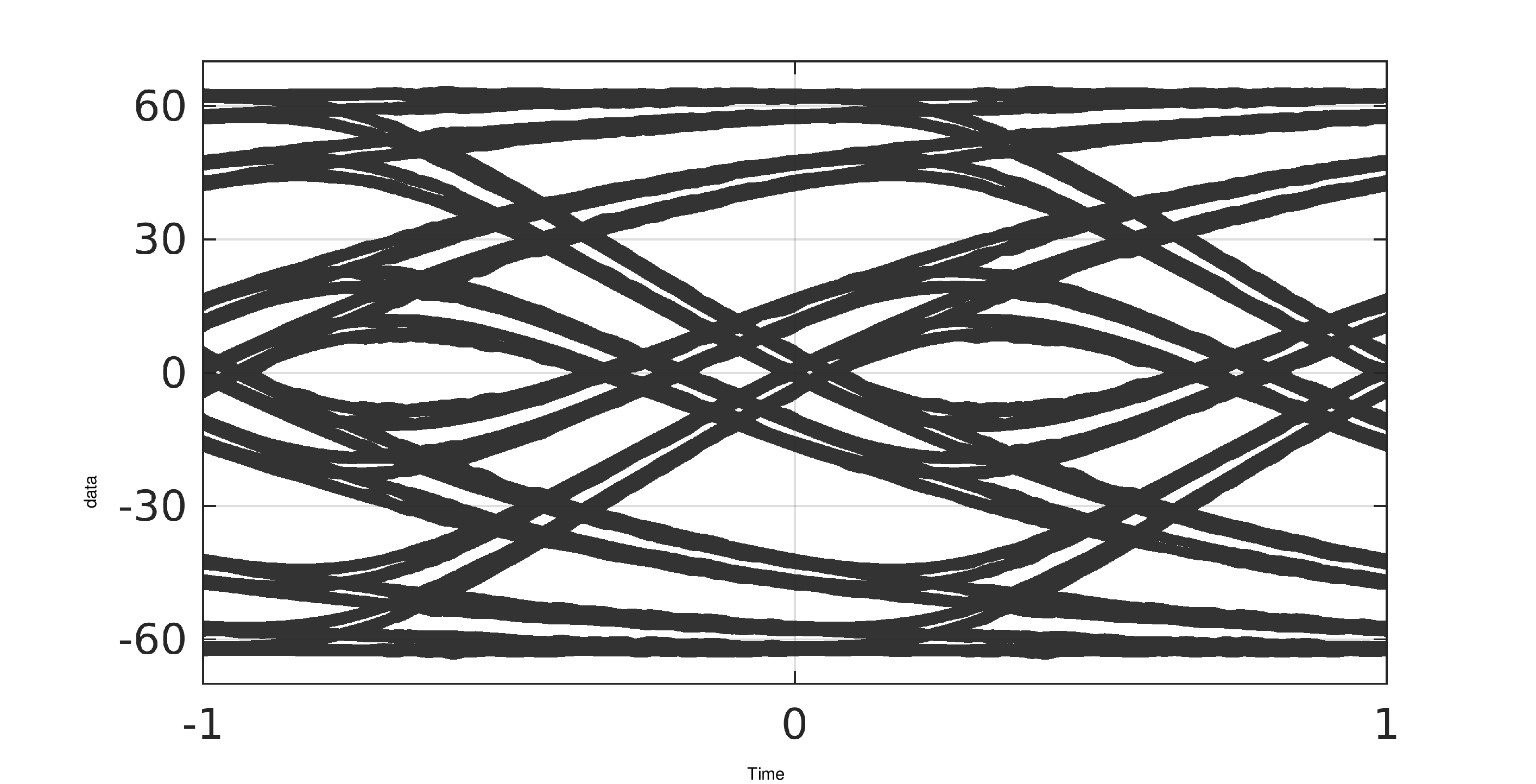}
\caption{Receiver input eye diagram for a channel with high ISI.}
\label{fig:ch6:coarse-eye-high}
\end{figure}

\begin{figure}[h!]
\centering
\psfrag{Time (mus)}{\scriptsize{Time ($\mu$s)}}
\psfrag{Coarse only}{\scriptsize{\hspace{-2ex}Coarse enable}}
\psfrag{Voltage}{\scriptsize{\hspace{-6ex}Control voltage (V)}}
\includegraphics[width=0.45\columnwidth]{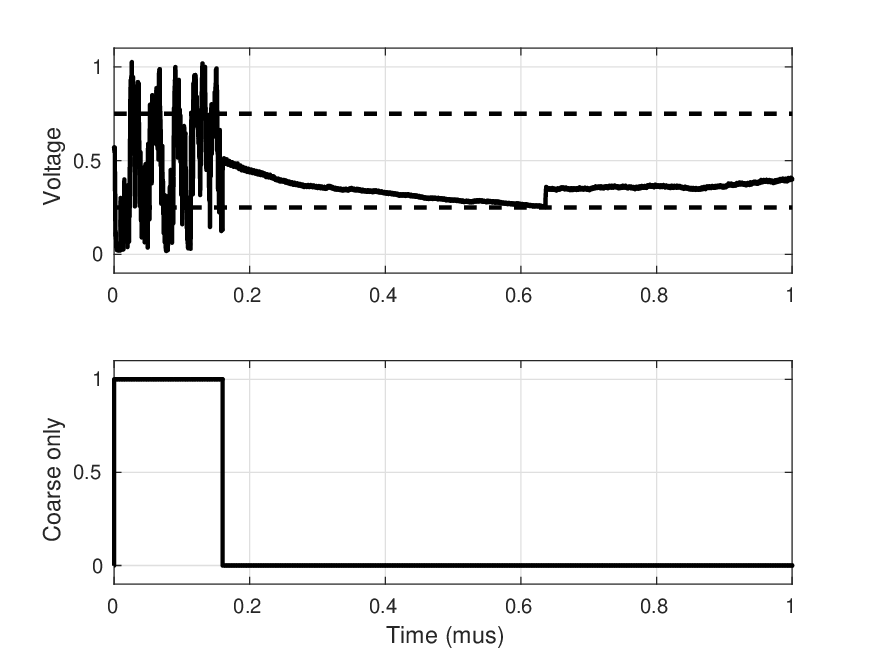}
\caption{Settling of the coarse first synchronizer when the input 
signal has high ISI with only 50\% horizontal eye opening. The lower 
waveform is the control signal for  enabling the fine correction.}
\label{fig:ch6:coarse-first-highisi}
\end{figure}

This same synchronizer circuit was simulated with an input that had 
very high \gls{isi}. The receiver input eye diagram for this test is 
shown in Fig.~\ref{fig:ch6:coarse-eye-high}.
With such data input, the settling of the synchronizer is shown in 
Fig.~\ref{fig:ch6:coarse-first-highisi}. 
The clock was initialized to the center of the window $\cW$.
Even for this input, the circuit escapes the window $\cW$ within the 
coarse first operating period. The noise in the control voltage is 
primarily due to the high jitter in the incoming data.

\subsection{Synchronizer with a bias towards one of the absorbing states}
%Variance of the time to absorption and effect of
%UP and DN mismatch}
%As discussed in the previous section, the settling time of 
%mesochronous clock retiming circuits can be quite high along with a large 
%standard deviation. In this section, 
Reduction of settling time using the coarse first technique is specific 
to the coarse+fine type synchronizer in~\cite{naveen_vlsi17}.
A different, though more general, technique of reducing the settling 
time is by introducing a mismatch in the relative strengths of the UP 
and DN updates in every cycle. 
%is discussed. 
%In this technique, a deliberate relative mismatch in the 
%strengths of the UP and DN updates is introduced.  
Using the 1 bit ISI model for the purpose of illustration, 
we next discuss the settling time of this synchronizer.

\begin{figure}[h!]
\centering
\psfrag{Tw/2}{\scriptsize{$T_{\cW}/2$}}
\psfrag{-Tw/2}{\scriptsize{-$T_{\cW}/2$}}
\psfrag{Mean absorbtion time (#cycles)}{\small{\hspace{3ex}Mean absorption 
time (cycles)}}
\psfrag{Initial clock phase}{\small{Initial clock position}}
\psfrag{No mismatch}{\scriptsize{No mismatch}}
\psfrag{10\% mismatch}{\scriptsize{10\% mismatch}}
%\subfloat[Mean absorption time]{
%\includegraphics[width=\columnwidth]{../figs/mean_mismatch_plot}
%\label{fig:ch6:markovmodel-mean-mismatch}}
\psfrag{Tw/2}{\small{$T_{\cW}/2$}}
\psfrag{-Tw/2}{\small{-$T_{\cW}/2$}}
\psfrag{Standard deviation (#cycles)}{\small{\hspace{-2ex}Standard deviation (cycles)}}
\psfrag{Initial clock phase}{\small{Initial clock position}}
\psfrag{No mismatch}{\scriptsize{No mismatch}}
\psfrag{10\% mismatch}{\scriptsize{10\% mismatch}}
\psfrag{Tw/2}{\scriptsize{$T_{\cW}/2$}}
\psfrag{-Tw/2}{\scriptsize{-$T_{\cW}/2$}}
%\subfloat[Mean absorption time]{
\includegraphics[width=0.5\textwidth]{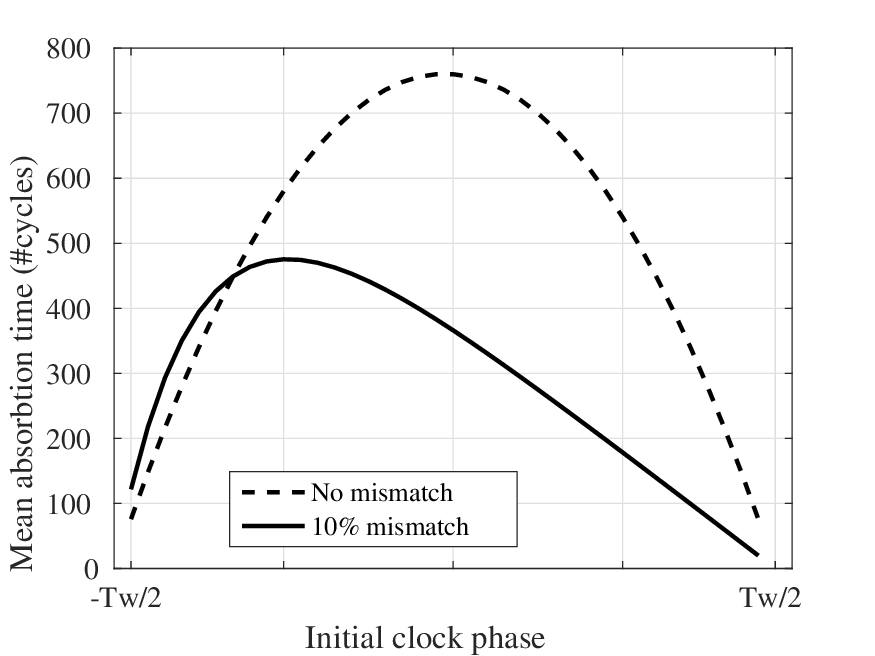}
%\label{fig:ch6:markovmodel-mean-mismatch}}
%\subfloat[Standard deviation of the absorption time]{
%\includegraphics[width=0.9\columnwidth]{../figs/variance_plot}
%\label{fig:ch6:markovmodel-std-mismatch}}
\caption{Mean absorption time 
%and (b) standard deviation of the 
for a symmetric Markov chain and for a chain with a 
10\% bias to right shift.}
\label{fig:ch6:markovmodel-mismatch}
\end{figure}

\begin{figure}[h!]
\centering
\psfrag{Tw/2}{\scriptsize{$T_{\cW}/2$}}
\psfrag{-Tw/2}{\scriptsize{-$T_{\cW}/2$}}
\psfrag{Mean absorbtion time (#cycles)}{\small{\hspace{-4ex}Mean absorption 
time (cycles)}}
\psfrag{Initial clock phase}{\small{Initial clock position}}
\psfrag{No mismatch}{\scriptsize{No mismatch}}
\psfrag{10\% mismatch}{\scriptsize{10\% mismatch}}
%\subfloat[Mean absorption time]{
%\includegraphics[width=\columnwidth]{../figs/mean_mismatch_plot}
%\label{fig:ch6:markovmodel-mean-mismatch}}
\psfrag{Tw/2}{\small{$T_{\cW}/2$}}
\psfrag{-Tw/2}{\small{-$T_{\cW}/2$}}
\psfrag{Standard deviation (#cycles)}{\small{\hspace{3ex}Standard deviation (cycles)}}
\psfrag{Initial clock phase}{\small{Initial clock position}}
\psfrag{No mismatch}{\scriptsize{No mismatch}}
\psfrag{10\% mismatch}{\scriptsize{10\% mismatch}}
\psfrag{Tw/2}{\scriptsize{$T_{\cW}/2$}}
\psfrag{-Tw/2}{\scriptsize{-$T_{\cW}/2$}}
%\subfloat[Mean absorption time]{
\includegraphics[width=0.5\textwidth]{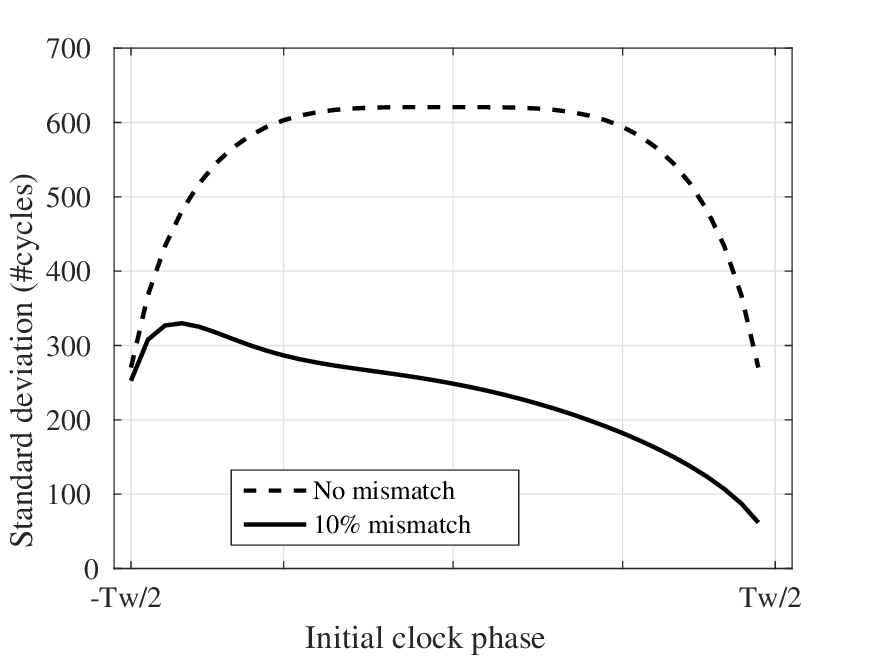}
%\label{fig:ch6:markovmodel-mean-mismatch}}
%\subfloat[Standard deviation of the absorption time]{
%\includegraphics[width=0.9\columnwidth]{../figs/variance_plot}
%\label{fig:ch6:markovmodel-std-mismatch}}
\caption{Standard deviation of the 
for a symmetric Markov chain and for a chain with a 
10\% bias to right shift.}
\label{fig:ch6:markovmodel-mismatch-std}
\end{figure}

Fig.~\ref{fig:ch6:markovmodel-mismatch}
shows the mean time to absorption as
a function of initial sampling phase when the step sizes for 
the left and the right shifts are (i) equal, and (ii) mismatched by 10\%.
The standard deviation for these cases is give in 
Fig.~\ref{fig:ch6:markovmodel-mismatch-std}.

Note that a 10\% mismatch in the step size of the left/right shift 
reduces the mean absorption time by up to 40\%. The effect of this 
mismatch is more pronounced in the standard deviation of the absorption 
time as a small mismatch is shown to result in a large reduction in the 
standard deviation as shown in 
Fig.~\ref{fig:ch6:markovmodel-mismatch-std}.
Hence to reduce the settling time of the circuit, 
one can deliberately introduce a mismatch in the UP/DN strengths
of the charge pump in the circuit. Alternately, a training sequence 
can also induce a similar mismatch, by producing different amounts of 
ISI to the two logic levels 0 and 1. This inherently introduces a 
bias towards one side of the window $\cW$ and reduces the settling time. 

\subsubsection{Reduction in settling time with charge pump mismatch}
\begin{figure}[h!]
\centering
\psfrag{Control voltage}{\small{Control voltage (V)}}
\psfrag{Time (mus)}{\small{Time ($\mu s$)}}
\psfrag{Without CP mismatch}{\footnotesize{Without CP mismatch}}
\psfrag{With a 10\% CP mismatch}{\footnotesize{With 10\% CP mismatch}}
\includegraphics[width=0.45\columnwidth]{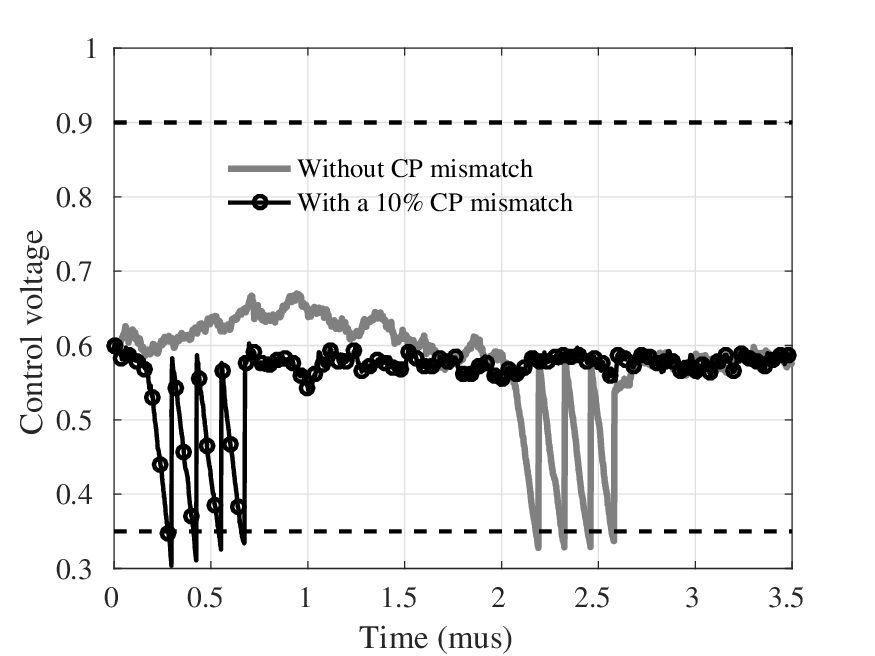}
\caption{Control voltage of the clock recovery circuit when 
$\Delta \phi \sim \pi$ radians, showing the quick escape from the window $\cW$ 
when the charge pump~(CP) has a mismatch in the UP and DN currents.}
\label{fig:ch6:vc-cp-mismatch}
\end{figure}

Fig.~\ref{fig:ch6:vc-cp-mismatch} shows the reduction in the settling 
time of the circuit with the introduction of a 10\% mismatch in the UP 
and the DN currents of the charge pump. Under identical initial 
conditions, a simulation with 10\% mismatch showed $>$80\% reduction 
in the time taken to exit the window $\cW$.

Charge pump mismatch, however, can result in increased jitter once the 
circuit has locked. 
\textcolor{blue}{Simulations of the coarse+fine retiming circuit show 
that a 10\% mismatch in the values of the UP and DN currents increases 
the peak-to-peak jitter of the recovered clock from 2.04~ps to 
2.27~ps, which is approximately 
a 10\% increase in the jitter. This is still only about 0.6\% of the 
clock period of 400~ps which was used in this simulation. 
If such a performance degradation is not acceptable, 
%Hence if good jitter performance is desired, 
the introduced mismatch can be switched off after lock has been 
achieved.}
%if good jitter performance is desired.

\subsubsection{Reduction in settling time with training sequence}
A training sequence that biases the system to either one of the 
directions can be used to reduce the settling time. For instance, 
one such training sequence is ``...0010011100100111...". This sequence 
has a minimum run length of 1 bit for a logic `1' and 2 bits for logic
`0'. Hence the ISI for logic `1' is always more than that for 
logic `0' and the phase detector's output in the window $\cW$
is biased towards the left. \textcolor{blue}{For instance, assuming 
the ISI is limited to 1 previous bit, we can use 
Table~\ref{tbl:ch6:isi_1bit} to compute the probabilities of 
a corrective action to the left and to the right respectively. 
In the above bit pattern, we have the following 3 bit sequences:
a) ``001'', b) ``010'', c) ``100'', d) ``001'' e) ``011'', f) ``111'', 
g) ``110'', h) ``100'', which then repeat. Hence, referring to 
Table~\ref{tbl:ch6:isi_1bit}, we see that $\prob(NA) = 0.5$, 
$\prob(LT) = 3/8$ and $\prob(RT) = 1/8$. Hence, this sequence 
implements a 25\% bias to the left.}

\begin{figure}[h!]
\centering
\psfrag{Control voltage}{\small{Control voltage (V)}}
\psfrag{Time (mus)}{\small{Time ($\mu s$)}}
\psfrag{Random data}{\footnotesize{With random data}}
\psfrag{Training data}{\footnotesize{With training data}}
\includegraphics[width=0.45\columnwidth]{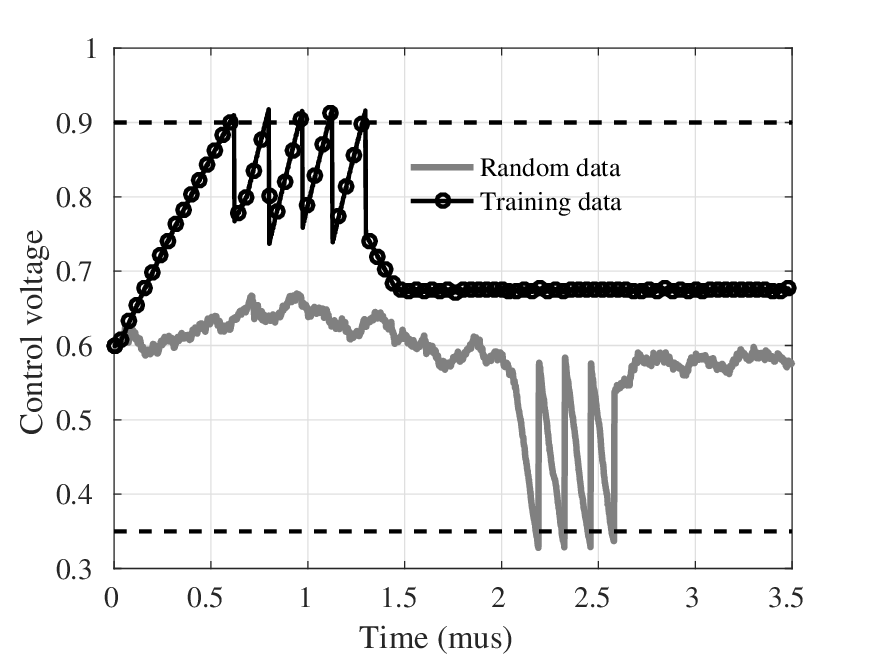}
\caption{Control voltage of the clock recovery circuit when 
$\Delta \phi \sim \pi$ radians, showing the quick escape from the window $\cW$
when a biased training sequence is used.}
\label{fig:ch6:vc_with_pilot}
\end{figure}

Fig.~\ref{fig:ch6:vc_with_pilot} shows the 
control voltage as the circuit settles when (a) the above training 
sequence ``...0010011100100111..." is used, and (b) a random 
equiprobable binary sequence is used.
Notice the considerable reduction in the settling time when the 
deliberately biased training sequence is used. One could also use an
alternating 1 and 0 training sequence which reduces the jitter due to 
ISI to zero. However, random uncorrelated jitter between the 
transmitter and receiver clocks may still result in a non-zero size 
of the window $\cW$.

%% file: conclusions.tex
\section{Conclusions}
\label{sec:conclude}
The effect of jitter on the settling time of mesochronous
clock retiming circuits is discussed in detail. In particular, 
it is shown how ISI 
induced jitter and random jitter can increase the settling time of 
clock recovery circuits indefinitely. A model of the system as a Markov 
chain with absorbing states is developed. 
Using this model, the effect of different types of jitter is analyzed.
%These include, data dependent jitter introduced by 1 previous bit, by 2 previous 
%bits, random jitter, and simultaneous data dependent and random jitter. 
The model predictions of the settling time in terms of the mean 
absorption time of the Markov chain match well with behavioural 
simulations. Using insights gained from the model, techniques for 
reducing the settling time are reported.
A coarse first synchronizer that uses only coarse correction steps 
initially is proposed. This architecture achieves small settling 
times in the presence of substantial 
\gls{isi}. Another technique of reducing the settling time, by 
introducing a mismatch between the phase updates in either direction 
of the clock retiming circuit, is also discussed. This is applicable to 
phase interpolator based clock retiming circuits as well.
%and also to the clock  retiming circuit discussed in Chapter~\ref{ch4:cdr}. 
This mismatch 
is achieved either by introducing a mismatch in the charge pump 
or through biased training data.
%or by using appropriately designed training data. 
All the suggested 
fast settling synchronizers are verified with circuit simulations.

%\begin{acknowledgments}
%The authors would like to thank Nagendra Krishnapura and Shanthi Pavan, 
%both from IIT Madras, for providing access to the equipment at 
%the VLSI Testing lab of IIT Madras for testing the fabricated chip.
%\end{acknowledgments}

%% file: appendix.tex
\section*{Appendices}
\appendix
\section{Derivation of the deterministic component of the settling time}
\label{apdx:ts}
For mesochronous systems, a clock running at the correct frequency is 
available and only the phase has to be corrected.
Hence, the clock recovery circuit can be of the first 
order, i.e., the loop filter is a single 
capacitor~\cite{naveen_vlsi17,phase_interpolator}.
%Hence, the 
%loop filter is a single capacitor. The settling time is derived in 
When a bang-bang phase detector is used, the
capacitor voltage ($V_c$) is quantized to a step size given by
\begin{align*}
\delta V_c = \int_0^T \frac{I_{CP}}{C} dt = \frac{K_{CP}}{C}.
\end{align*}
Here $K_{CP}$ is the gain of the charge pump, expressed in terms of the 
charge pump current $I_{CP}$ and the clock period $T$. 
Hence, the step size of the phase corrections is 
\begin{align}
\delta \phi = \delta V_c K_{VC} = \frac{K_{CP}K_{VC}}{C}.
\label{eqn:ch6:phi_step}
\end{align}
The number of phase correction
steps ($M$) needed for achieving lock can be written as 
\begin{align} 
M=
\begin{cases}
\frac{\Delta \phi}{\delta \phi} & \mbox{ when } \Delta \phi < \pi, \\
\frac{2\pi - \Delta \phi}{\delta \phi} & \mbox{ when } \Delta \phi > \pi .
\end{cases}
\label{eqn:ch6:phierror}
\end{align}
The settling time can then be written as 
\begin{align}
t_s=\frac{MT}{\alpha},
\label{eqn:ch6:ts}
\end{align}
where \gls{dact} is the data activity factor.
From \eqref{eqn:ch6:phi_step},~\eqref{eqn:ch6:phierror} and \eqref{eqn:ch6:ts}, 
the settling time expression in~\eqref{eqn:ch6:ts_deterministic} follows.
%
%\begin{align}
%%\label{eqn:ch6:ts_deterministic}
%t_s&=T\frac{C}{\alpha K_{VC}K_{CP}} \Delta \phi &\mbox{ when } \Delta  \phi < \pi , \nonumber \\ 
%t_s&=T\frac{C}{\alpha K_{VC}K_{CP}}(2\pi - \Delta \phi) &\mbox{ when } \Delta  \phi > \pi .
%\end{align}

\section{Mean time to absorption of a Markov chain}
\label{apdx:markov-meantime}
Knowing the probability transition matrix of a Markov chain, the mean 
time to absorption and its variance can be computed. We will outline an 
example computation in this Appendix.
The state diagram of the Markov chain representation of the clock recovery 
circuit, for data with 1 bit ISI, is shown 
in Fig.~\ref{fig:ch6:mc_model}. The states corresponding to $-T_{\cW}/2$ 
and $T_{\cW}/2$, which are 
at the edges of the window of susceptibility, are absorbing states. 
%
\begin{comment}
\begin{figure}[h]
\centering
\psfrag{p0}{\small{$\frac{1}{2}$}}
\psfrag{p+}{\small{$\frac{1}{4}$}}
\psfrag{p-}{\small{$\frac{1}{4}$}}
\psfrag{k}{\scriptsize{$T_w$/2}}
\psfrag{-k}{\scriptsize{$-T_w/2$}}
\psfrag{1}{\scriptsize{1}}
\psfrag{0}{\scriptsize{0}}
\includegraphics[width=8.5cm]{../figs/MC2}
\caption{\small Absorbing Markov chain model of the synchronizer}
\label{fig:mc_model1}
\end{figure}
\end{comment}
%
The probability transition matrix $\bar{P}$ for this Markov 
chain can be written as 
\begin{align*}
\bar{P} = 
 \begin{bmatrix}
%State & 0 & 1 & 2 & \cdots & k-2 & k-1 & k \cr
  1 & 0 & 0 & 0 & \cdots & & & & 0 \cr
  \frac{1}{4} & \frac{1}{2} & \frac{1}{4} & 0 & \cdots & & &  & 0 \cr
   0 & \frac{1}{4} & \frac{1}{2}  & \frac{1}{4} & \cdots & &  & & 0 \cr
   & \vdots & &  & \ddots & & & \vdots &  \cr
   0 & \cdots & & & & 0 & \frac{1}{4} & \frac{1}{2} & \frac{1}{4} \cr
   0 &  \cdots & & & & 0 & 0 & 0 & 1 \cr
  \end{bmatrix}.
\end{align*}
To calculate the mean time to absorption (and the variance of the time 
to absorption), $\bar{P}$ is first written in the canonical form
\cite{markov}.  This is obtained by reordering the entries 
in $\bar{P}$ to separately aggregate all the transient and 
absorbing states respectively.
Hence, $\bar{P}$ can be written as 
%\begin{IEEEeqnarray*}{rCl}
\begin{align*}
\bar{P}&=\left[
\arraycolsep=2.5pt\def\arraystretch{1.5}
\begin{array}{c:c}
Q & R \\ \hdashline
0 & I
\end{array}\right] \\
%
%\begin{bmatrix}[c|c]
%	Q & R \cr \hline
%	0 & I_t \cr
%\end{bmatrix}
&=
% \begin{bmatrix}[ccccccc|cc]
\left[
\arraycolsep=2.5pt\def\arraystretch{1.5}
\begin{array}{ccccccc:cc}
  \frac{1}{2} & \frac{1}{4} & 0 & 0 & \cdots & & 0 & \frac{1}{4} & 0 \\
   \frac{1}{4} & \frac{1}{2} & \frac{1}{4} & 0 & \cdots & & 0 & 0 & 0 \\
    & \vdots & & & \ddots & & \vdots & \vdots & \vdots \\
   0 & \cdots & & & & \frac{1}{4} & \frac{1}{2} & 0 & \frac{1}{4} \\ \hdashline
   0 & \cdots & & & & 0 & 0 & 1 & 0 \\
   0 &  \cdots & & & & 0 & 0 & 0 & 1 \\
  \end{array}
\right].
%\end{IEEEeqnarray*}
\end{align*}
Here $Q$ is a square matrix which captures the transitions from one 
transient state to another transient state, $R$ captures the transitions 
from transient states to absorbing states and $I$ is an identity matrix. The fundamental matrix $N$ can be computed using the equation
\begin{equation*}
N= (\tilde{I} - Q)^{-1},
\end{equation*}
where $\tilde{I}$ is an identity matrix of the same dimensions as $Q$.
Conditioned on the starting position, the mean time to absorption 
($T_{mean}$), in terms of the number of steps, is given by
\begin{align*}
 T_{mean} = NC^1,
\end{align*}
where $C^1$ is a column vector with number of rows equal to the number 
of transient states, and with all entries as 1. The variance of the 
time to absorption (in terms of number of steps), conditioned on 
the starting position, is calculated as
\begin{align*}
 T_{var} = (2N-\tilde{I})\cdot T_{mean} - T_{mean}^{sq}.
\end{align*}
Here $T_{mean}^{sq}$ is obtained by squaring the elements of $T_{mean}$.
%$N$ is the fundamental matrix computed as $(I_t - Q)^{-1}$.

%% file: bio.tex
\begin{IEEEbiography}
[{\includegraphics[width=1in,height=1.25in,clip,keepaspectratio]{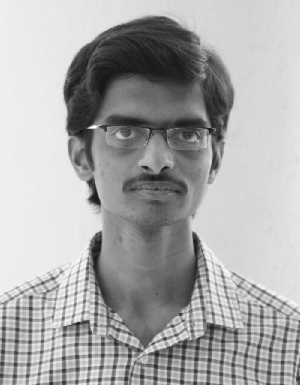}}]{Naveen Kadayinti}
received his M.Tech + Ph.D. dual degree from the Indian Institute of 
Technology Bombay in 2017. He is currently an Assistant Professor 
in the Department of Electrical Engineering at the Indian Institute of Technology 
Dharwad. 
He was a visiting faculty member in the Department of Avionics 
at the Indian Institute of Space Science and 
Technology Trivandrum during Spring 2017-18.

His research interests include wired and wireless communication circuits
and mixed signal SoC design and test.
\end{IEEEbiography}

\begin{IEEEbiography}   
[{\includegraphics[width=1in,height=1.25in,clip,keepaspectratio]{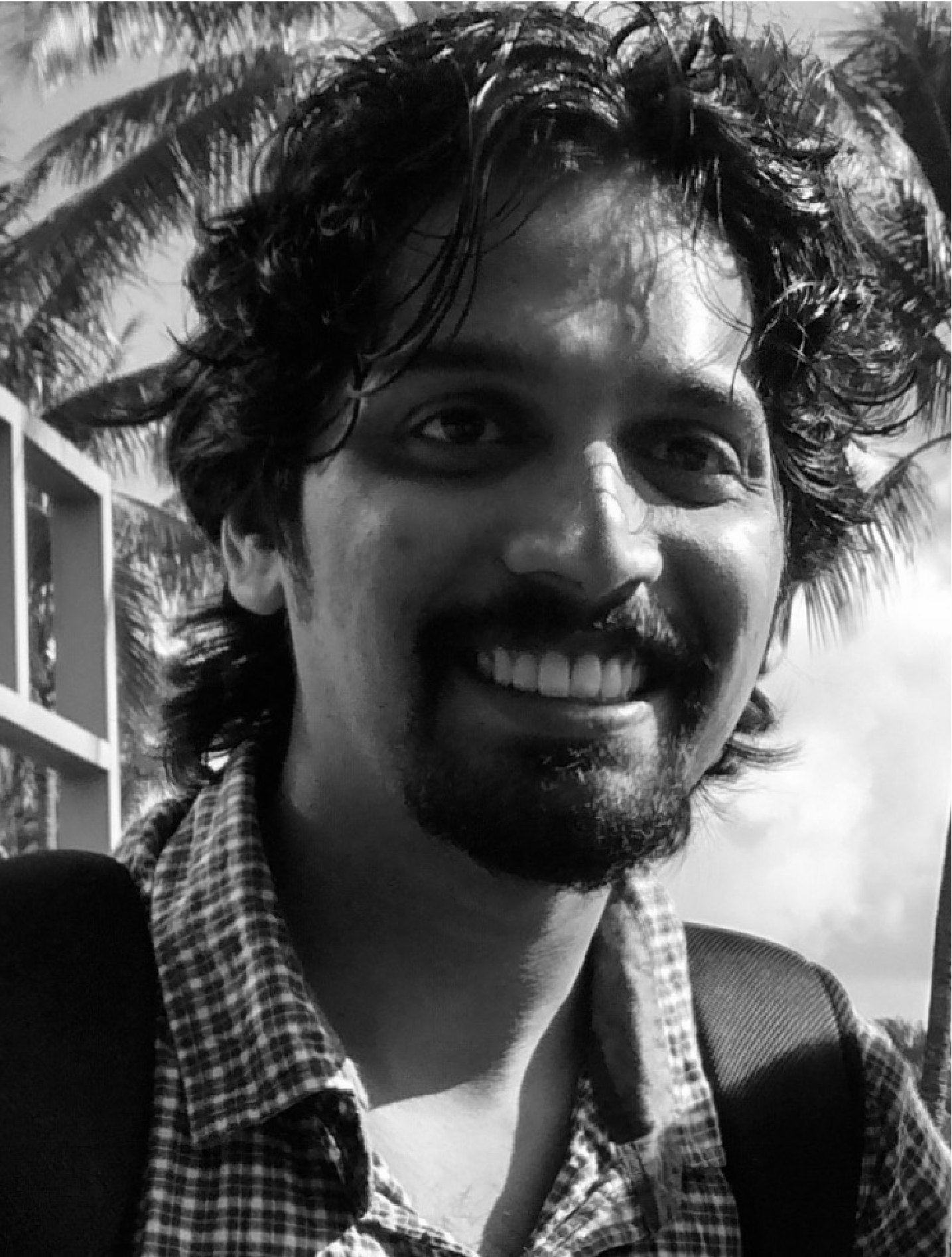}}]{Amitalok J. Budkuley} 
received the B.E. degree in Electronics and Telecommunications 
Engineering from Goa University, Goa, India in 2007, and the M.Tech. 
and Ph.D. degree in Electrical Engineering from the Indian Institute 
of Technology Bombay, Mumbai, India in 2009 and 2017 respectively. 
From 2009 to 2010, he was with Cisco Systems, Bangalore, India. 
He is currently a Postdoctoral Researcher at the Department of 
Information Engineering, The Chinese University of Hong Kong, 
Sha Tin, Hong Kong. 

His research interests include information theory, communication 
systems and game theory.
\end{IEEEbiography}

\begin{IEEEbiography}[{\includegraphics[width=1in,height=1.25in,clip,keepaspectratio]{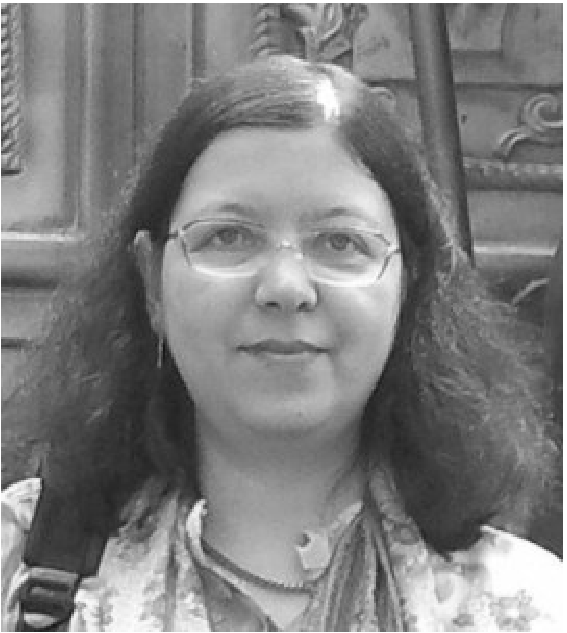}}]{Maryam Shojaei Baghini}
(M'00 - SM'09) received the M.S. and Ph.D. degrees
in electrical engineering from Sharif University of Technology, Tehran, in
1991 and 1999, respectively. She has worked for more than 2
years in industry on the design of analog ICs. In 2001, she joined
Department of Electrical Engineering, IIT-Bombay, as a Postdoctoral
Fellow, where she is currently a Professor. She is the author/coauthor of
143 international journal and conference papers, the
inventor/coinventor of 6 granted US patents, 1 granted Indian patent and
25 more filed patent applications.  Her current research interests include
high-speed links for on-chip and off-chip data communication,
high-performance low-energy analog/mixed-signal/RF IC design and test for
various applications including healthcare and sensor networks,
device-circuit co-design in emerging technologies and energy harvesting
circuits and systems. Dr. Shojaei serves in the Technical Program
Committee of several IEEE conferences, including IEEE International
Conference on VLSI Design, and Asia Symposium on Quality Electronic
Design. She was a TPC member of IEEE-ASSC from 2009 to 2014. Dr. Shojaei
is joint recipient of 11 awards
and a senior member of IEEE.
\end{IEEEbiography}

\begin{IEEEbiography}[{\includegraphics[width=1in,height=1.25in,clip,keepaspectratio]{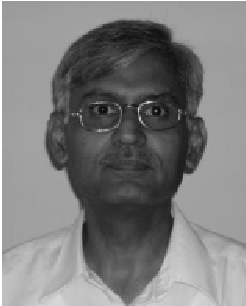}}]{Dinesh Sharma} 
obtained his Ph.D. from the Tata Institute of Fundamental
Research, Mumbai. He has worked at TIFR and IITB at Mumbai, 
at LETI at Grenoble in France and at the Microelectronics Center of North 
Carolina in the U.S.A. on MOS technology, devices and mixed mode circuit 
design. He has been at the EE deptt. of IIT Bombay since 1991, where he 
is currently a Professor. His current interests include mixed
signal design, interconnect technology and the impact of technology on 
design styles.

He is a senior member of IEEE, a fellow of IETE and has served on the 
editorial board of "Pramana", published by the Indian Academy of Science.
\end{IEEEbiography}